\documentclass[]{article}
\usepackage{graphicx}
\usepackage{natbib}
\usepackage{bm}
\usepackage{epsfig}
\usepackage{inputenc}
\usepackage{color}
\usepackage{array}
\usepackage{pstricks}
\usepackage{amsmath,amssymb,mathrsfs}
\usepackage{pict2e}
\usepackage{supertabular}
\usepackage{soul}
\usepackage[a4paper,pdftex,verbose,margin=3.0cm,twoside]{geometry}
\let\realverbatim=\verbatim
\let\realendverbatim=\endverbatim
\renewcommand\verbatim{\par\addvspace{6pt plus 2pt minus 1pt}\realverbatim}
\renewcommand\endverbatim{\realendverbatim\addvspace{6pt plus 2pt minus 1pt}}
\makeatletter
\newcommand\verbsize{\@setfontsize\verbsize{10}\@xiipt}
\renewcommand\verbatim@font{\verbsize\normalfont\ttfamily}
\makeatother

\newcommand{\bfvec}[1]{\mbox{\boldmath{$#1$}}}

\newcounter{hours}\newcounter{minutes}

\newlength{\CheckBarLenA}
\newlength{\CheckBarLenB}
\newlength{\CheckBarLenC}

\def\eqalign#1{\null\,\vcenter{\openup\jot\m@th
\ialign{\strut\hfil$\displaystyle{##}$&
$\displaystyle{{}##}$\hfil\crcr #1\crcr }}\,}

\def\é{\'e}
\def\è{\`e}
\def\à{\`a}
\def\ù{\`u}
\def\ê{\^e}
\def\ç{\cc}
\def\â{\^a}
\def\î{\^i}
\def\û{\^u}
\def\ô{\^o}
\def\ï{\i}
\def\vk{\bfvec k}

\def\vu{\bfvec u}
\def\ve{\bfvec e}

\def\v0{\bfvec 0}

\def\vN{\bfvec N}

\newcommand{\TS}{\mathsfbf{S}}

\def\I{{\cal I}}

\def\I{\uppercase\expandafter{\romannumeral 1}}
\def\II{\uppercase\expandafter{\romannumeral 2}}
\def\III{\uppercase\expandafter{\romannumeral 3}}
\def\IV{\uppercase\expandafter{\romannumeral 4}}

\def\Rh{\hat R}

\def\uh{\hat u}

\def\vuh{\hat{\vu}}

\def\vuh{\hat{\vu}}

\def\ltextindent#1{\hbox to \hangindent{#1\hss}\ignorespaces}

\newlength{\FigureHeight}
\newlength{\FigureHeightHalf}

\def\k{\kappa}

\def\vp{\mbox{\boldmath$p$}}
\def\vq{\mbox{\boldmath$q$}}

\DeclareMathAlphabet{\mathsfbf}{OT1}{cmss}{bx}{n}
\DeclareMathAlphabet{\mathmibf}{OT1}{cmr}{bx}{it}

\newsavebox{\astrutbox}
\sbox{\astrutbox}{\rule[-5pt]{0pt}{20pt}}

\newenvironment{DIFnomarkup}{}{}

\title{Quasi-static magnetohydrodynamic turbulence at high Reynolds number}

\author{B. Favier$^{1,3}$, F.S. Godeferd$^1$\footnote{Corresponding author: Fabien.Godeferd@ec-lyon.fr}, C. Cambon$^1$, A. Delache$^2$, W.J.T. Bos$^1$ \\
$^1$ LMFA UMR 5509 CNRS, \'Ecole Centrale de Lyon, Universit\'e de Lyon, France.\\
$^2$ LMFA@UJM St-\'Etienne, CNRS UMR 5509 \\
Universit\'e de St-\'Etienne, F-42023 Saint-\'Etienne Cedex 2, France. \\
$^3$ School of Mathematics and Statistics, Newcastle University, UK.}
\date{}
\begin{document}

\label{firstpage}
\maketitle

\begin{abstract}
We analyse the anisotropy of homogeneous turbulence in an electrically conducting fluid 
submitted to a uniform magnetic field, for low magnetic Reynolds number, 
in the quasi-static approximation. We interpret  disagreeing previous predictions
between linearized theory and simulations:
in the linear limit, 
the kinetic energy of transverse velocity components, 
normal to the 
magnetic field, decays faster than the kinetic energy of the axial 
component, along 
the magnetic field (\cite{moff67}); whereas
 many numerical studies predict a final state characterised 
by dominant energy of transverse velocity
components.
We investigate the corresponding nonlinear phenomenon using Direct Numerical 
Simulations  of freely-decaying turbulence, and a two-point statistical spectral closure 
based on the Eddy Damped Quasi-Normal 
Markovian model.
The transition from the three-dimensional turbulent flow 
 to a ``two-and-a-half-dimensional'' 
flow (\cite{mont82}) is a result of the combined effects of
short-time linear Joule dissipation  and longer time
nonlinear creation of polarisation anisotropy.
%WB
It is this combination of linear and nonlinear
effects which explains the disagreement between predictions
from linearized theory and results from numerical simulations. The transition
is characterized by 
%WB
the elongation of turbulent structures along
the applied magnetic field, and by the strong anisotropy
of directional two-point correlation spectra, in agreement
with experimental evidence. Inertial equatorial transfers in both DNS
and the model are presented  to  describe in detail the most important equilibrium dynamics.
Spectral scalings are maintained
in high Reynolds number turbulence attainable only with the EDQNM
model, which also provides simplified modelling of the
asymptotic state of quasi-static MHD turbulence.
%\textcolor{red}{Abstract \`a reprendre substantiellement.}
\end{abstract}

%\begin{keywords}
Magnetohydrodynamics, Quasi-static hypothesis, Homogeneous turbulence, Direct Numerical Simulations, EDQNM
%\end{keywords}
 
%\tableofcontents
\section{Introduction}
%----------------------
In most geophysical and astrophysical flows, turbulence is affected by forces
that distort significantly some of its scales in an anisotropic manner, such
as the Coriolis force in rotating flows or the Lorentz force arising from 
the presence of an external magnetic field
in a conducting fluid.
This specific turbulent dynamics forced by an imposed magnetic field is
 found in liquid
metal flows, be they of industrial, geophysical nature---the melted iron 
core of the earth---or
of academic interest in the laboratory, such as the
experiment by \cite{ALEMANY-MOREAU-SULEM-FRISCH} in liquid mercury.
Recent laboratory experiments on the
dynamics of conducting fluids use sodium or  gallium;  liquid sodium
 is also used in industrial configurations, for instance in the French fast breeder 
reactor Superph\'enix.

Generally, the motion of turbulent liquid metals is governed by
magnetohydrodynamics (MHD): the induction equation for the fluctuating
magnetic field is added to the Navier-Stokes equations,
which are in turn modified by the Lorentz force,
representing the feedback from the magnetic field.
In the presence of an external
magnetic field, such MHD coupling results
in new dissipative terms, of ohmic nature, and selectively
damped waves, the Alfv\'en waves (\cite{moff67}).
In cases involving liquid metal, the magnetic diffusivity in the 
induction equation is larger than the  molecular
viscosity in the Navier-Stokes equations, \textit{i.e.} the magnetic 
Prandtl number is small compared to one.
The magnetic diffusivity is so large with respect to the
kinematic diffusivity---with a magnetic Prandtl number less
than $10^{-5}$ in the Earth's iron core, of order
$0.9\times 10^{-5}$ in liquid sodium or $1.4\times 10^{-7}$ in mercury---that it is consistent
to consider the flow at very high  Reynolds number and
at low magnetic Reynolds number. In the following simulations,
the magnetic Prandtl number is set to $\textit{Pr}_M=3.1\times 10^{-4}$.

As discussed in section \ref{sec:equation}, if the 
magnetic Reynolds number
is small enough, the linear regime no longer admits 
Alfv\'en waves solutions,
and the effect of the Lorentz force reduces to an anisotropic
ohmic (or Joule) dissipation term. In this regime, called the 
quasi-static approximation (QS MHD), the induction equation is  
simple enough to be solved explicitly and to yield 
a closed expression of the Lorentz force in terms of the velocity.
The specificity of the quasi-static limit can be discussed both in terms of
timescales and anisotropy. Unlike more general MHD turbulent flows,
in which nonlinear and Alfv\'en timescales may be in competition
and yield length scale dependent levels
of anisotropy (see \textit{e.g.} \cite{ZHOU-MATTHAEUS, ZHOU10}),
in QS MHD the magnetic diffusivity is too large to enable
Alfv\'en waves.
%field is 
%a mere by-product of the velocity field, with timescale of its own.
The only relevant timescales concern the modified Navier-Stokes equations, 
with a
linear nondimensional timescale $\eta/B_0^2$ resulting from 
ohmic dissipation ($B_0$ is the external magnetic field, scaled as velocity, and $\eta$
the magnetic diffusivity), and the nonlinear timescale 
$l_0/u_0$ ($u_0$ is the \textit{rms} velocity
and $l_0$ the length scale related to a turnover time).
 Strong anisotropy
is first induced by the ohmic dissipation term over the linear timescale. 
% In contrast, a larger
% number of timescales are called into play in `full' MHD, as discussed,
% e.g. by Zhou \& Matthaeus (Phys. plasmas 2005), Zhou et al. (review, 2004),
% and some aspects of Alv\'enic MHD can be understood in an `isotropized'
% formalism (Kraichnan ? Ironishkov ?).

%%% ajout de paragraphe
%+
Quasi-static MHD turbulence was investigated experimentally
by \cite{ALEMANY-MOREAU-SULEM-FRISCH} and \cite{CAPERAN-ALEMANY}.
In these studies, turbulence was generated by towing a grid
through a cylindrical tank full of mercury, with
%\marginpar{\textcolor{red}{ donner des$\rightarrow$ d\'etails sur les r\'esultats
%spectraux}}
an external magnetic field generated by 
a coil.
Measurements include Reynolds stress components, an integral length
scale in the axial direction and one-dimensional spectrum
of transverse energy with respect to the axial wavenumber. A clear
transition from a three-dimensional state, with conventional Kolmogorov spectrum, to
a quasi-two-dimensional state, with $k^{-3}_{\parallel}$ spectrum, was evidenced.
%% In addition to Alemany {\it et al.} (1979), already mentioned, the
%% experimental results
%% were reported only in French papers, such as in Cap\'eran \& Al\'emany,
%% 1985, and in Garnier {\it et al.} 1981.
The first phase of this 3D-2D transition was studied using
axisymmetric Lin equations with
an Eddy Damping Quasi-Normal Markovian (EDQNM) closure model  
by \cite{CAMBON-1990}, and the scenario of a
two-dimensionalization in two steps was proposed. This scenario
was recently confirmed by DNS in \cite{favier10} and one of
the goals of the present paper is to go  beyond the numerical approach
of \cite{CAMBON-1990} using both anisotropic EDQNM and direct numerical simulations
(DNS). The ``eddy-damping'' rate appearing in the EDQNM closure for general MHD 
turbulence should in principle be modified to account for
the combination of sweeping and straining mechanisms, thus allowing
for the possibility of either Kolmogorov inertial scaling ($k^{-5/3}$ kinetic
energy spectra, isotropized, \textit{i.e.} spherically integrated) or 
Iroshnikov-Kraichnan scaling ($k^{-3/2}$) (see \textit{e.g.} \cite{ZHOU-MATTHAEUS-DMITRUK}). 
The QS MHD approximation, without Alfvenic propagation,
 allows to  anchor the model within the classical 
hydrodynamic turbulence context, thus keeping the original damping
consistent with Kolmogorov scaling.

Other numerical
approaches in the same context are given by \cite{schu76}, and \cite{knae04}
with application to anisotropic modelling. A survey is offered by
\cite{knae08}, in which the change of anisotropic
structure for the Reynolds stress tensor, from purely linear to
nonlinear dynamics, is presented as an open problem. We think that
this problem can be elucidated by the scenario of 3D-2D transition in two
steps
(\cite{CAMBON-1990, favier10}) which is fully described hereafter.

Both spectral theory and DNS were applied by \cite{ISHIDA-KANEDA}
to the dynamical and
structural study of the small scales anisotropy of
QS MHD turbulence, while a
 recent approach by \cite{okamoto_2010} focused on
the infrared limit, \textit{i.e.} at very large scales. In the latter
work, assuming the
%\begin{verbatim}
%\marginpar{\textcolor{red}{ il faudrait$\rightarrow$ fournir quelques d\'etails sur leurs r\'esultats...une phrase ou deux. Claude ?}}
%\end{verbatim}
existence of a Loitsyanski-like invariant, decay laws for typical
integral lengthscales and  Reynolds stress components are proposed
and compared to DNS results. 
The dynamics of  integral length scales was shown to be crucial in
rotating turbulence which bears strong analogies with MHD turbulence. 
For instance, the \textit{linear}  growth rate 
of the integral length scale related to transverse velocity components
and axial separation, denoted $\ell_{\parallel}$ in 
\cite{okamoto_2010}, and $L^{(3)}_{11} = L^{(3)}_{22}$ here, was clearly
related to the  role of \textit{nonlinear} transfer terms 
(\cite{camb89,jacq90,CAMBON-MANSOUR-GODEFERD}). This result was recently
recovered by \cite{staplehurst} with a different interpretation,
although we believe that  
the use of axisymmetric Lin equations---equations
for two-point velocity correlation spectra (\cite{VONKARMAN-LIN})---, in which
linear and nonlinear terms are exactly separated, is essential to
the understanding. Accordingly, our theoretical approach is based
on an anisotropic spectral formalism with generalized Lin equations instead 
of on a
formalism based on the
K\`arm\`an-Howarth equation, rather used by \cite{okamoto_2010}, but
bridges between the two approaches will be discussed in the following.
%+
%%% fin paragraphe ajouté

One of the most challenging aspects of quasi-static MHD turbulence, 
from a numerical point of view, 
is the rapid increase of the velocity correlation lengths in 
the direction of the imposed magnetic field.
In that case, the results from classical pseudo-spectral methods with 
periodic boundary condition are often questionable, as the characteristic 
scale of the turbulent motion is no longer small compared to the numerical 
box size.
In this paper, we  compare Direct Numerical Simulations 
(DNS) with a model based on EDQNM
closures and confirm that neither the low Reynolds numbers considered in 
DNS nor the confinement due to periodic boundary conditions alter our 
understanding of the dynamics.  
Secondly, our goal is to propose a detailed study of the anisotropy of 
quasi-static MHD turbulence at low, moderate and high Reynolds numbers. 
As in \cite{favier10}, the analogy with the asymptotic quasi-two-dimensional 
state, called ``two-and-a-half-dimensional'' flow, will also be discussed.

The paper is organised as follows.
The main parameters and governing equations are recalled in the following 
section.
Spectral properties and EDQNM closures are discussed in 
section \ref{sec:exact}, 
and the numerical methods used in the paper are presented in 
section \ref{sec:numerical}.
Section \ref{sec:conf} is devoted to the issue of confinement, 
both in DNS and EDQNM.
Most of the results are gathered in section \ref{sec:comp}, where the 
statistical properties of quasi-static MHD turbulence are described, with
an emphasis on anisotropy characterisation (section~\ref{sec:aniso}).
Finally,  the large Reynolds number behaviour is investigated in 
section \ref{sec:edqnmres}, along with the analogy with quasi-two-dimensional 
turbulence in section \ref{sec:2D3C}.
Details about EDQNM closed equations and linear predictions for the 
velocity correlation lengths are gathered in 
Appendices~\ref{appendixA} and~\ref{appendixB}.
%
%%%%%%%%%%%%%%%%%%%%%%%%%%%%%%%%%%%%%%%%%%%%%%%%%%%%%%%%%%%%%%%%%%%%%%%%%%%%%%%%%%%%
%
\section{Governing equations and parameters}
%-----------------------------------------------
\label{sec:equation}
We consider initially isotropic homogeneous turbulence in an incompressible 
conducting fluid, in which $u_x \simeq u_y \simeq u_z$, where $u_x$, $u_y$
and $u_z$ are the \textit{rms} values of the velocity components.
When the external magnetic field is applied, along $z$ in the following, $u_z$
will be called the \textit{axial} component and $u_x$, $u_y$ the \textit{transverse}
components. 
%are the horizontal components of the velocity whereas $u_z$ is the 
%vertical one.
The fluid is characterised by the kinematic viscosity $\nu$,  
density $\rho$ and  magnetic diffusivity $\eta=(\sigma\mu_0)^{-1}$; 
$\sigma$ is the electrical conductivity, 
$\mu_0$ the magnetic permeability.
These physical properties are assumed to be constant.
%The initial \textit{rms} velocity is $u_0$ and t
The integral length scale is $l_0$, defined from the
two-point velocity 
correlation tensor $R_{ii}(r)=\langle u_i(x_i)u_i(x_i+r)\rangle$, 
as $l_0=\int_0^\infty R_{ii}(r)/R_{ii}(0) \mathrm{d} r$, (or equivalently
from the kinetic energy spectrum).
The Reynolds number and its magnetic counterpart are 
$\textit{Re}=(u_0l_0)/\nu\gg1$ and $R_M=(u_0l_0)/\eta\ll1$.
The ratio between these two numbers defines the magnetic Prandtl number 
$\textit{Pr}_M=\nu/\eta$, which is very small in our study. 
The flow is submitted to a uniform vertical magnetic field $\bm{B}$ 
scaled as Alfv\'en speed as $\bm{B}_0=\bm{B}/\sqrt{\rho\mu_0}$.
The ratio between the eddy turnover time $l_0/u_0$ and the ohmic time 
$\eta/B_0^2$ is the magnetic interaction number $N=(B_0^2l_0)/(\eta u_0)$.
Within the quasi-static approximation, which implies that $R_M$ tends to zero, 
but which is nonetheless approximately valid for all $R_M<1$ (\cite{knae04}), 
the Navier-Stokes equations become
%\vspace{-1mm}
%
\begin{equation}
\label{eq:momentum}
\frac{\partial\bm{u}}{\partial t}+\bm{u}\cdot\nabla\bm{u}
 =-\frac{1}{\rho}\nabla p+\nu\nabla^2\bm{u}+\underbrace{M_0^2\Delta^{-1}
    \frac{\partial^2\bm{u}}{\partial z^2}}_{\bm{F}}
\end{equation}
where $\bm{F}$ is the rotational part of the Lorentz force, $\Delta^{-1}$ 
is the inverse of the Laplacian operator, $M_0^2=B_0^2/\eta$ and $z$ the 
axial coordinate, along the direction of $\bm{B}_0$. Compressible effects 
are not taken into account here, so that $\nabla\cdot\bm{u}=0$.
%
%%%%%%%%%%%%%%%%%%%%%%%%%%%%%%%%%%%%%%%%%%%%%%%%%%%%%%%%%%%%%%%%%%%%%%%%%%%%%%%%%%%%%%
%
\section{Exact and model equations for two-point second-order statistics}
%-------------------------------------------------------
\label{sec:exact}
We obtain hereafter the equations for the spectral
statistics of the second-order moment
of the fluctuating velocity field $\bm{u}$.
The derivation is facilitated in two ways: first, by beginning with
the Fourier coefficients of $\bm{u}$ before computing
the second-order moments; second, by using a Helmholtz-like decomposition
 in order to derive all the algebra only in terms of the
incompressible components, namelly the toroidal/poloidal decomposition.

Equation~(\ref{eq:momentum}) for the velocity is 3D-Fourier transformed, with
Fourier coefficients denoted with $\widehat{\hspace*{1em}}$,
and the pressure term is eliminated using incompressibility, 
introducing Kraichnan's projector 
\begin{equation}
P_{imn}(\vk) = - \frac{\mathrm{i}}{2} \left[k_m \left(\delta_{in} - \frac{k_ik_n}{k^2}\right) 
+ k_n\left(\delta_{im} - \frac{k_i k_m}{k^2}\right)\right] \ ,
\end{equation}
so that
\begin{equation}
\left(\frac{\partial }{\partial t} + \nu k^2 + M_0^2 \cos^2 \theta \right)
\uh_i (\vk,t) = P_{imn}(\vk) \widehat{u_m u_n},
\label{utfourier} 
\end{equation} 
where $\vk$ is the wave vector and $\theta$ its orientation with respect to the $z$-axis.
The unique new term reflecting the quasi-static MHD effect is algebraic, 
$\left(M_0^2\cos^2\theta\right)\uh_i(\vk, t)$.

For second-order velocity correlations, the most general information is given 
by the second-order spectral tensor $\Rh_{ij}(\vk, t)$ which in the homogeneous 
case is given by
\begin{equation}
\langle \uh^*_j(\vp, t) \uh_i(\vk, t) \rangle = \Rh_{ij}(\vk, t) \delta^3 (\vk - \vp) \ .
\label{eq:tensrij}
\end{equation}
The 3D Dirac function expresses that only the Fourier velocity components at 
the same wave vector have non zero double correlation.
Another expression is obtained by considering a discretized velocity 
field, as in DNS 
%and integrating the domain over $\vp$ 
(thus turning
the mathematical formalism of distributions and generalized integrals,
applied in continuous space, to 
classical integrals applied to discretized functions).
For the particular case of a cubic periodic domain of size $L$, 
this replaces the Dirac term in the above equation by a factor $(L/(2\pi))^3$.

The brackets in equation~(\ref{eq:tensrij}) denote statistical ensemble averaging: in
DNS started with a single realization of the velocity field, statistical averaging
is obtained by spatial averaging, assuming ergodicity and using
the particular symmetries preserved here, namely axisymmetry.

\begin{figure}
\unitlength 0.5mm
\begin{picture}(300,120)
        \put(80,0){\includegraphics[height=110\unitlength]{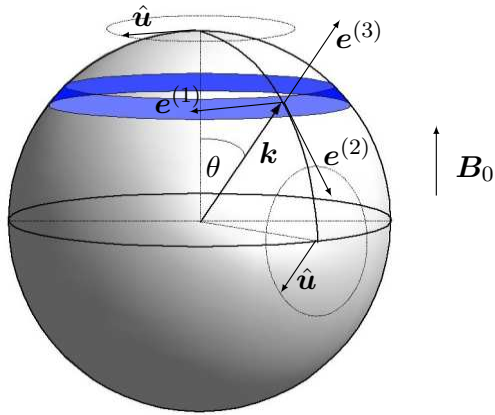}}
\end{picture}
\caption{\label{fig:craya}Craya-Herring frame 
$(\bm{e}^{(1)},\bm{e}^{(2)},\bm{e}^{(3)})$ in Fourier space.  
In the general case, Fourier modes 
in the blue region contribute to  $E(k,\theta)$ (eq.\eqref{eq:ekt_def2}).
However, if $\vk$ is vertical, the sum of the two components $u^{(1)}(\vk)$
and $u^{(2)}(\vk)$
 generates a vertically sheared 
horizontal flow (VSHF), and  if $\vk$ is horizontal,
  they correspond to transverse and axial 
components.
Therefore, the polar 
modes ($\theta\simeq 0$) contribute to horizontal 
kinetic energy, whereas equatorial modes ($\theta\simeq\pi/2$) 
contribute to both axial (along $\bm{e}^{(2)}$) and transverse 
(along $\bm{e}^{(1)}$) \textcolor{black}{kinetic} energies.}
\end{figure}

In the quasi-static MHD case under consideration, statistical symmetry 
is thus restricted to axisymmetry with mirror symmetry 
(\textcolor{black}{the mean helicity 
is zero} if initially zero), and the spectral tensor can be expressed in terms of toroidal 
and poloidal components of the velocity field 
in Fourier space.
The two components are obtained 
 using a polar-spherical frame of reference with base vectors
$\ve^{(1)}(\vk)$ and $\ve^{(2)}(\vk)$
 (\textit{a.k.a.} 
Craya-Herring frame of reference, see figure \ref{fig:craya}; \cite{herr74}), as 
\begin{equation}
\vuh (\vk, t) = u^{(1)}(\vk, t) \ve^{(1)}(\vk) 
     + u^{(2)}(\vk, t)\ve^{(2)}(\vk) \ . \label{eq:crayaherring}
\end{equation}
%or the basis of helical modes, as
%\begin{equation}
%\vuh(\vk, t) = \xi_+ (\vk, t) \vN(\vk) + \xi_- (\vk, t) \vN^*(\vk)\ .
%\end{equation}
%The base vectors in the two decompositions are simply related through 
%\begin{equation}
%\vN(\vk) = \ve^{(2)}(\vk) - \textrm{i} \ve^{(1)}(\vk), \quad 
%\vN^* (\vk) = \vN(-\vk) \ .
%\end{equation}
This decomposition automatically 
treats the velocity field as solenoidal, \textit{i.e.}
 divergence free in physical space, 
through the algebraic  orthogonality condition  
$\bm{k}\cdot\hat{\bm{u}}(\bm{k}) =0$. In addition,
it allows to construct any related statistical correlation,
with a minimal number of components, for arbitrary anisotropy.
The decomposition~(\ref{eq:crayaherring}) is general,
although, since it relies on the arbitrary choice of a polar axis,
it is especially well suited to axisymmetric configurations, 
in which the tensors' dependence reduces to the
wavenumber $k$ and its angle $\theta$ to the axis. 

%The directions $1$ and $2$ are interpreted in physical space as 
%toroidal and poloidal components (e.g. \cite{saga08}).
%

The expression for the two-point second-order spectral 
tensor is therefore
\begin{equation}
\label{eq:poltor}
\Rh_{ij} = \Phi^1 e^{(1)}_i e^{(1)}_j + \Phi^{2} e^{(2)}_i e^{(2)}_j \ ,
\end{equation}
in which all the tensors and vectors depend on $k$ and $\theta$ as the
toroidal and poloidal energy tensors
\begin{equation}
\Phi^1 (\vk,t) = \Phi^1 (k, \cos\theta,t), \quad \Phi^2 (\vk, t) = \Phi^2(k, \cos\theta,t) \ .
\end{equation} 
Considering the symmetries of the flow, 
the most general decomposition in terms of energy density $e$, 
polarization $\mathcal{Z}$ and helicity $\mathcal{H}$ reduces to (\cite{camb89})
\begin{equation}
e(k, \cos\theta, t) = \frac{1}{2} \left( \Phi^1 + \Phi^2 \right), \quad \mathcal{Z}(k, 
        \cos\theta, t) = 
       \frac{1}{2} \left( \Phi^2 - \Phi^1 \right), \quad {\cal H}(\vk, t) =0 \ . \label{eq:eZH}
\end{equation}
The polarization term $\mathcal{Z}$ is in general complex-valued  and 
its imaginary part corresponds to a non zero cross-correlation between 
poloidal and toroidal velocity components.
Here, $\mathcal{Z}$ is real-valued and both sets of statistical quantities, 
$\Phi^1$ and $\Phi^2$, or $e$ and $\mathcal{Z}$ are equivalent.

It is straightforward 
to derive the following exact equations for 
%$\Rh_{ij}$, or for 
$e$ and $\mathcal{Z}$:
\begin{eqnarray}
\left(\frac{\partial }{\partial t} + 2 \nu k^2 + 2M_0^2 \cos^2 \theta \right)
e(k, \cos\theta, t) = T^{(e)}(k, \theta, t) \label{line}\\
\left(\frac{\partial }{\partial t} + 2 \nu k^2 + 2M_0^2 \cos^2 \theta \right)
\mathcal{Z}(k, \cos\theta, t) = T^{(\mathcal{Z})}(k, \theta, t). \label{linz}
\end{eqnarray}
These equations are exact in the limit of homogeneous quasi-static 
MHD turbulence.
They generalise the Lin equation, with the definition of cubic 
$T^{(e,\mathcal{Z})}$ terms given in \cite{camb89}, and recalled in 
appendix~\ref{appendixA}.
All the terms in these equations can be obtained in pseudo-spectral DNS, 
as in \cite{favier10}, using summation of Fourier modes on rings,
in contrast with the summation of Fourier modes
on spherical shells as usual in the analysis of isotropic turbulence
(see figure~\ref{fig:craya}).
However, the anisotropic $(k, \theta)$ 
distribution of $T^{(e,\mathcal{Z})}$ 
is more affected by lack of sampling and noise in DNS, 
especially at small $k$ where $\Delta k/k$ is large. 
It is therefore worthwhile
to develop a model based on equations~(\ref{line}) and~(\ref{linz})
to evaluate the behaviour of the second- and third-order moments---energy and
energy transfer spectra. The model may then provide  smooth values for these 
quantities, to be quantitatively compared to DNS results.

We will be using hereafter such a model, drawn from
the anisotropic EDQNM closure theory, which has already been successfully 
applied to rotating or stably stratified turbulent flows, including
a comparison with DNS 
(\cite{CAMBON-MANSOUR-GODEFERD,GODEFERD-STAQUET}).
%
%Statistical quantities of particular interest are the Reynolds Stress components and length scales based on both velocity and vorticity.
%A splitting in terms of $e$ and $\mathcal{Z}$ is readily obtained for all these
%quantities, resulting in a link to structure-based modeling (Kassinos 
%{\it et al.} 2001, and references herein) and various angles (e.g.
%Shebalin 1993). Definition of these quantities are recalled in section 7.
%
%%%%%%%%%%%%%%%%%%%%%%%%%%%%%%%%%%%%%%%%%%%%%%%%%%%%%%%%%%%%%%%%%%%%%%%%%%%%%%%%%%%%
%
%\section{Anisotropic EDQNM closures based on helical mode decomposition}
%-------------------------------------------------
%\label{sec:aniso}
%
In the derivation of the model,
the toroidal/poloidal decomposition proves useful and valuable 
for simplifying the 
expressions for triple velocity correlations, without using projection
operators inherited from equation (\ref{utfourier}). 
Another simplification comes from the use of a slightly modified
decomposition of velocity, 
$\vuh(\vk, t) = \xi_+ (\vk, t) \vN(\vk) + \xi_- (\vk, t) \vN^*(\vk)$ 
analogous to~(\ref{eq:crayaherring}), 
which brings out the helical modes $\xi_\pm$ by projection onto
$\vN(\vk) = \ve^{(2)}(\vk) - \textrm{i} \ve^{(1)}(\vk)$ and $\vN^* (\vk) = \vN(-\vk)$.
Helical modes 
are advantageous because they diagonalise the curl operator
and allow a more compact decomposition of triple velocity correlations
at three points (triadic terms),
even in isotropic turbulence (see for example \cite{wale92}). 
[In rotating turbulence, the helical modes
are also the inertial waves modes, 
\cite{camb89,wale93,CAMBON-MANSOUR-GODEFERD,bell06}].
The starting point of the closure is the third-order spectral tensor 
$\TS$
related to
helical modes, defined by
\begin{equation}
\langle \xi_{s''} (\vq,t) \xi_{s'}(\vp,t) \xi_s(\vk, t) \rangle 
= S_{s s' s''} (\vk, \vp, t) \delta^3 (\vk + \vp + \vq).
\end{equation}
The generalised Eddy Damping Quasi-Normal (EDQN) technique is then applied to the equation 
that governs the third-order spectral tensor,
\begin{multline}
\Big[ \frac{\partial}{\partial t} + \nu\left(k^2 + p^2 + q^2\right)\\
+ M_0^2 \left(\cos^2 \theta_k + \cos^2 \theta_p + \cos^2 \theta_q\right)\Big] 
S_{s s' s''}(\vk, \vp,t) = \Omega_{s s' s''} (\vk, \vp, t),
\label{cubict}
\end{multline}
in which $\Omega_{s s' s''}(\vk, \vp, t)$ represents the contribution
of fourth-order velocity correlations. 
In order to obtain a closed set of equations, 
$\mathsfbf{\Omega}$ is  expressed in terms of sums of products of
double correlations. This would be an  exact evaluation
of the fourth-order moments, were it applied to a Gaussian random variable
(the `QN' part).
We apply 
a corrective term (the `ED' part) due to the non-vanishing fourth-order cumulant,
to account for the departure from Gaussianity of both
third-order and fourth-order cumulants. 
%Details of the EDQNM model and
%of the different versions we will be discussing in the following
%are provided in appendix~\ref{appendixA}.
We shall use the version of the model that
has provided the best results in rotating or stably stratified turbulence.
This EDQNM2 model, say, accounts for the anisotropic
Joule dissipation in both the second-order moments equation and in
the third-order moments one~(\ref{cubict}).
When informative, the results of EDQNM2 will also be contrasted with those of
the simpler EDQNM1 model, which retains the Joule dissipation
term only in the second-order moment equation, discarding it in equation~(\ref{cubict}). 
Contrasting both
models allows to tell whether the main anisotropic mechanism is mostly
linear or nonlinear. Additional 
information on the models is given in appendix~\ref{appendixA}.

%
%%%%%%%%%%%%%%%%%%%%%%%%%%%%%%%%%%%%%%%%%%%%%%%%%%%%%%%%%%%%%%%%%%%%%%%%%%%%%%%%%%
%
\section{Numerical methods}
%----------------------------
\label{sec:numerical}
To assess the validity of EDQNM closure in the context of quasi-static 
MHD turbulence and to obtain results at low and moderate Reynolds numbers, 
we perform Direct Numerical Simulations of equation \eqref{eq:momentum} using 
a pseudo-spectral method implemented on a parallel computer. 
The velocity field is 
computed in a cubic box of 
side $L$ with periodic boundary conditions using $512^3$ Fourier modes. 
[The conventional shorthand relationship  $L=2\pi$  for non-dimensional DNS
is used here, except for the previous discussion 
after equation~(\ref{eq:tensrij}).]
A spherical $2/3$-truncation of Fourier modes is used 
to avoid aliasing and the time scheme is third-order Adams-Bashforth. 
The dissipative viscous plus ohmic terms  are treated implicitly.

The DNS results presented here are performed at higher resolution
than those of \cite{favier10}.
An initially isotropic turbulent velocity field is created by
 a  hydrodynamic simulation  with  
large-scale forcing in order to reach a quasi-steady state.
At the end of this pre-computation stage, the \textit{rms} velocity 
is $u_0=0.81$ and the integral scale $l_0=0.25$ yielding 
$\textit{Re}=u_0l_0/\nu\simeq333$.
The Reynolds number based on the Taylor microscale is $R_{\lambda}\simeq95$.
This rather low value, considering the resolution, 
is a consequence of our specific choice of 
a small initial integral scale $l_0$, in order
to lift partially the  numerical confinement constraint, 
discussed in section \ref{sec:conf}. 
The corresponding turbulent flow field is used as initial state for 
two different MHD simulations.
In all of them $R_M\simeq0.1$ (hence $N\simeq2$), so that the 
%\textcolor{red}{Benjamin: c'\'etait bien
%$N$ plut\^ot que $\eta$ ? Il y avait une faute de frappe, je pense...}), so that the 
quasi-static approximation is justified (\cite{knae04}).
Two different amplitudes of the imposed magnetic field are chosen, 
which correspond to two values of the interaction parameter: $N=1$ and $5$.
For reference, we also compute the isotropic case, setting $B_0=0$, 
from the same initial condition.
The quasi-static MHD simulations are 
freely decaying to avoid spurious effects of a 
forcing scheme on the development of anisotropy.
\begin{figure}
\unitlength 0.5mm
\begin{picture}(300,110)
        \put(-5,0){\includegraphics[height=100\unitlength]{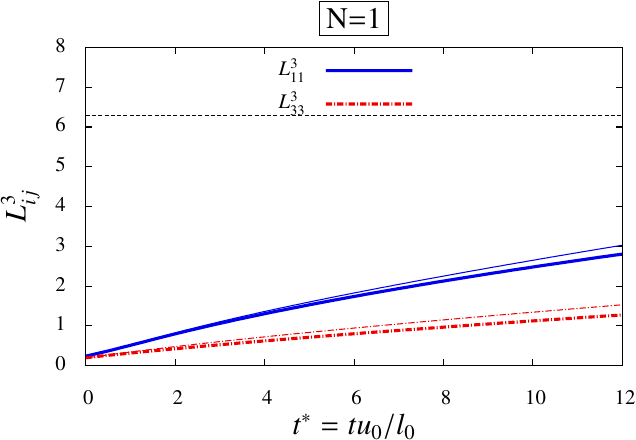}}
        \put(140,0){\includegraphics[height=100\unitlength]{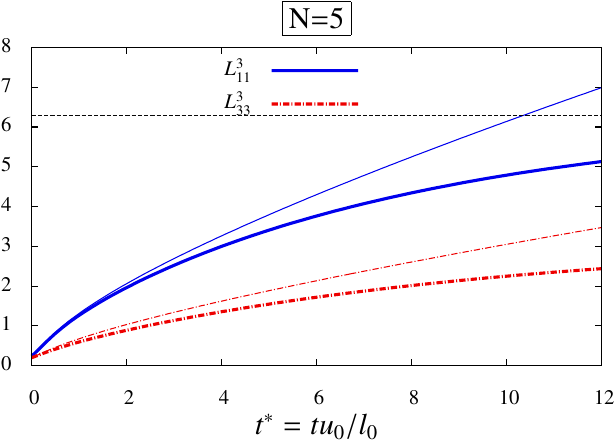}}
        \put(15,82){(a)}
        \put(148,82){(b)}
\end{picture}
\caption{Velocity correlation lengths versus dimensionless time $t^*=tu_0/l_0$. 
The thick lines correspond to DNS without nonlinear interactions. 
The thin lines correspond to analytical linear predictions from 
Appendix~\ref{appendixB}. 
The horizontal line presents the numerical limit of $2\pi$.\label{fig:lijlin}}
\end{figure}

The two versions EDQNM1 and EDQNM2 mentioned above are used.
%EDQNM3 was subject to numerical instability and was finally set aside.
The EDQNM simulations are initialised with the exact initial kinetic 
energy spectrum obtained from the DNS pre-computation.
%For the EDQNM computations presented in the following, 
The EDQNM spectral 
space is discretized as follows: we use $64$ values for the wave number $k$, 
$32$ for the polar angle $\theta$, 
and $32$ internal orientations for the angle defining the orientation of the plane
of the triads.
In contrast to DNS, the wave number discretization used here is logarithmic,
thereby improving the sampling of the large scales with
respect to DNS.
The minimum and maximum wave numbers solved are
$k_{\textrm{min}}=1$ and $k_{\textrm{max}}=512/3$ as in DNS.
If $B_0=0$, EDQNM1 and EDQNM2 are identical, and we also compute 
this particular case for comparison with isotropic DNS.
Hereafter, DNS results are plotted with lines only, 
 EDQNM results are plotted with lines and symbols ($\circ$ for 
EDQNM1 results, $\bullet$ for EDQNM2 results and $\triangle$ for isotropic EDQNM).
%
%%%%%%%%%%%%%%%%%%%%%%%%%%%%%%%%%%%%%%%%%%%%%%%%%%%%%%%%%%%%%%%%%%%%%%%%%%%%%%%%%%%%%%%%%%
%
\section{Confinement due to periodic boundary conditions}
%------------------------------------------------------------
\label{sec:conf}
This paragraph is specifically devoted to the problem of confinement 
in quasi-static MHD turbulence.
As the anisotropic ohmic dissipation affects the flow, 
the velocity field rapidly homogenizes in the direction of the 
imposed magnetic field.
The velocity correlation lengths thus increase in the axial direction.
However, due to periodic boundary conditions used in DNS, these correlation 
lengths are limited by the size  $L=2\pi$ of the computational domain.
To remove  possible non physical effects due to this
confinement, we compute the initial velocity field with an integral length 
scale about thirty times smaller than the numerical box size.
We therefore adopt an intermediate configuration with moderate value of 
the Reynolds number.

It is not possible to evaluate the finite-size effects
in the fully nonlinear case, especially because the theoretical study
is based on additional assumptions. So we will restrict our analysis
to the pure linear dynamics, or RDT. In so doing, we have to consider the following
caveat: the pseudo-spectral method is assumed to be ``exact'' in the linear limit---to a given accuracy
provided by the discretization in Fourier space---so that all RDT statistics derived
from averaging $\hat{u}_i^*\hat{u}_j$ cannot be directly affected by the finite-size effect. 
On the other hand, statistics calculated from velocity components in physical space may be affected,
even in the linear regime. 
%\textcolor{red}{FAB: ...hum...r\'efl\'echir...}

In order to assess the influence of the confinement
and the validity of DNS in the context of quasi-static 
MHD turbulence, we perform two simulations, for $N=1$ and $5$, in which  
the nonlinear advective term is neglected.
(Several comparisons of this type between linear and 
nonlinear simulations can be found 
in \cite{favier10}.)
These simulations can be compared to the linear analytical solutions from 
Rapid Distortion Theory (see \cite{moff67} and Appendix~\ref{appendixB}).
In order to study specifically the effect of confinement, 
we compute correlation lengths defined by
\begin{equation}
\label{eq:deflijl}
L_{ij}^{(l)}=\frac{1}{\left<u_iu_j\right>}\int_0^{\infty}
         \left<u_i(\bm{x})u_j(\bm{x}+\bm{r})\right>\textrm{d}\bm{r}
\end{equation}
where $r_k=r\delta_{kl}$ is the two-point velocity separation.
In the current axisymmetric flow, the most relevant anisotropy 
indicators are the integral length scales with axial separation,  
relative to either axial or transverse velocity components (\cite{camb89}):
\begin{align}
\label{eq:defl333}
L_{33}^{(3)} & =\frac{2\pi^2}{\langle u_3^2\rangle }\int_0^{\infty}
     \left[e(\bm{k})+\Re\mathcal{Z}(\bm{k})\right]\Big|_{k_z=0}k\textrm{d}k \\
\label{eq:defl113}
L_{11}^{(3)} & =\frac{\pi^2}{\langle u_1^2\rangle}\int_0^{\infty}
    \left[e(\bm{k})-\Re\mathcal{Z}(\bm{k})\right]\Big|_{k_z=0}k\textrm{d}k \ .
\end{align}
The expressions of linear solutions for
these quantities can be found in Appendix~\ref{appendixB}.
As discussed above, these quantities evaluated by DNS are expected to coincide
with these analytical formulas only  in the theoretical limit of a projection
base with an infinite number of degrees of freedom.
\begin{figure}
\unitlength 0.5mm
\begin{picture}(300,110)
        \put(-5,0){\includegraphics[height=100\unitlength]{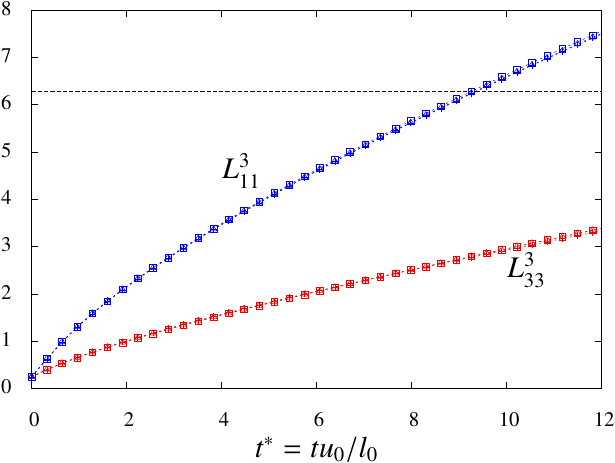}}
        \put(140,0){\includegraphics[height=100\unitlength]{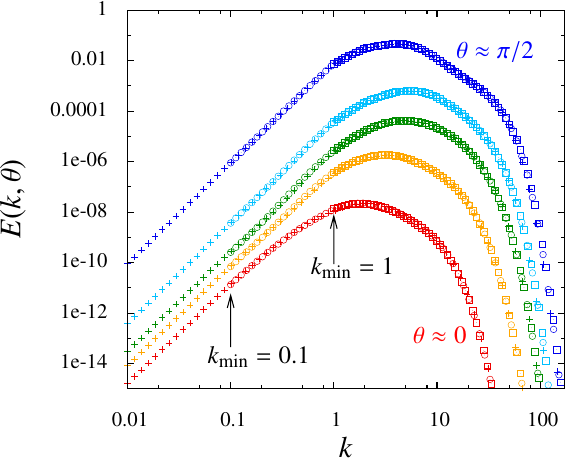}}
        \put(5,90){(a)}
        \put(170,90){(b)}
\end{picture}
\caption{Results from EDQNM2 with different minimum wave number, 
from $k_{\textrm{min}}=0.01$ to $k_{\textrm{min}}=1$. (a) Velocity correlation lengths. 
The horizontal dotted line corresponds to the numerical limit of $2\pi$, 
present in DNS. (b) Angular energy spectra at $t^*=12$.\label{fig:confedqnm}}
\end{figure}

Figure~\ref{fig:lijlin} shows the correlation lengths $L_{33}^{(3)}$ 
and $L_{11}^{(3)}$ versus the dimensionless time $t^*=tu_0/l_0$, starting
from the isotropic configuration at $t=0$.
The thick lines correspond to linearised DNS and thin lines 
correspond to linear analytical solutions.
Both $N=1$ and $N=5$ cases are presented, for which
one observes a growth of the correlation lengths, as expected in decaying
turbulence.
At moderate $N$, the length scales remain well below  
the numerical limit $L=2\pi$, although 
one still notices a small departure between DNS and RDT 
results. The length scale predicted by analytical RDT is
consistently larger than that of DNS. 
For $N=5$, one clearly observes that the 
vertical correlation length $L_{11}^{(3)}$ 
saturates before the maximum value $2\pi$ 
%(drawn as a thin dotted horizontal line in figure \ref{fig:lijlin}(b)), 
whereas the linear solution continues to 
grow. This difference is a clear example of confinement in anisotropic 
DNS and cannot be attributed to nonlinearities, which are absent in 
these simulations. 
% \begin{verbatim}
% It is also possible to compare results from linearised 
% DNS and analytical predictions concerning the polarization. 
% In the linear case, 
% $\mathcal{Z}=0$, so that $b_{33}^{\mathcal{Z}}=0$. 
% However, both $N=1$ and $N=5$ 
% DNS cases display a negative decreasing $b_{33}^{\mathcal{Z}}$ (not shown). 
% We therefore expect DNS to overestimate the absolute value for polarization 
% related quantities.
% \end{verbatim}

Note that the initial integral length scale in the present DNS is very small 
($l_0\approx0.25$) compared to the computational box size. 
Removing completely all trace of numerical confinement
would require decreasing $l_0$ even more. Considering the
current DNS resolution, the resulting Reynolds 
number would decrease too much for a turbulent flow to subsist. A 
solution is to increase the resolution, with increasingly 
demanding computational cost, to either a larger cubic box
with resolution $2048^3$, or an adapted elongated box 
with resolution $512^2\times 2048$,  as done by 
\cite{voro05} in MHD turbulence, or in
rotating turbulence by \cite{CAMBON-MANSOUR-GODEFERD} and in 
%stratified turbulence
%by \cite{BRETHOUWER-BILLANT-LINDBORG-CHOMAZ}.  
convective turbulence by \cite{MATSUMOTO}.
The latter
option indeed delays the confinement issue, which is most pregnant
in the axial direction, but also implies to some degree the anticipation
of the anisotropy in the later stage of the evolution.
 In the following, we retain a 
 $512^3$ resolution consistent with the isotropy of initial
conditions, considering only the early time response 
 $t^*\leq 6$, hence keeping the flow in a significantly nonlinear
regime while maintaining negligible confinement bias. 

In order to investigate whether the mechanisms observed at
these low Reynolds numbers will persist at higher Rynolds numbers, we will
use EDQNM closures.
%In this paper, we would like to demonstrate the possibility 
%to study quasi-static MHD turbulence using EDQNM closures.
It therefore makes sense to address also the problem of confinement in the
numerical resolution of EDQNM.
Such confinement limitations should in principle also apply to EDQNM models 
since the minimum wave number is, as in DNS, $k_{\textrm{min}}=1$. However, 
the closure model is written in spectral space so that periodic boundary 
conditions are not explicit.
The EDQNM spectral resolution can easily be increased in order to quantify the 
impact of numerical confinement through the value of the minimum resolved wave 
number $k_{\textrm{min}}$.
We thus perform three EDQNM2 simulations (the results are the same using 
EDQNM1) in the case $N=5$, with three different values 
  $k_{\textrm{min}}=0.01$, $0.1$, $1$.
%The results are gathered in figures \ref{fig:confedqnm}.
Firstly, the time evolution of the velocity correlation lengths 
$L_{33}^{(3)}$ 
and $L_{11}^{(3)}$ are plotted in figure \ref{fig:confedqnm}(a).
The predictions from the three simulations are 
almost undistinguishable, and, in contrast with the 
DNS results of figure \ref{fig:lijlin}(b), 
the growth of correlation lengths is not constrained by the value of the 
minimum wave number.
Secondly, angular energy spectra,
plotted on figure \ref{fig:confedqnm}(b) at $t^*=12$, show that
the spectral anisotropy is the same whatever $k_{\textrm{min}}$ 
%which is a confirmation that the 
%simulation with the largest value of $k_{\textrm{min}}$ is exempt of confinement. 
(details on the anisotropic spectra will be presented
in section \ref{sec:aniso}).
Accordingly, we choose $k_{\textrm{min}}=1$ in the following, 
to allow a complete comparison with DNS results, 
with the understanding that EDQNM is free from truncation effects.
\begin{figure}
\unitlength 0.5mm
\begin{picture}(300,110)
        \put(-5,0){\includegraphics[height=100\unitlength]{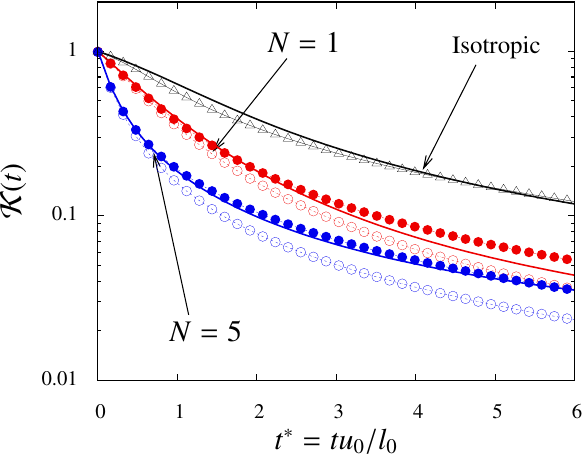}}
        \put(140,0){\includegraphics[height=100\unitlength]{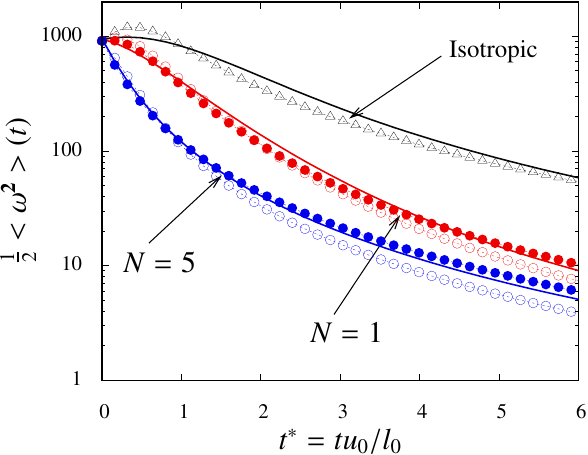}}
        \put(20,20){(a)}
        \put(184,20){(b)}
\end{picture}
\caption{\label{fig:kinetic-energy} Kinetic 
energy $\mathcal{K}(t)$ 
and enstrophy $<\bm{\omega}^2>/2$ 
versus dimensionless time $t^*$ in log-lin scale. --- DNS, $\circ$ EDQNM1, 
$\bullet$ EDQNM2, $\triangle$ isotropic 
EDQNM.}
\end{figure}

% \textcolor{red}{ALEX: si tu veux rajouter des \'el\'ements
% issus des \'echanges par mail de d\'ecembre, c'est dans ce paragraphe... Mais
% on a d\'ej\`a pas mail compl\'et\'e, \`a ce stade...}

%
%%%%%%%%%%%%%%%%%%%%%%%%%%%%%%%%%%%%%%%%%%%%%%%%%%%%%%%%%%%%%%%%%%%%%%%%%%%%%%
%
\section{Comparison between DNS and EDQNM}
%------------------------------------------
\label{sec:comp}
In this section, we propose a comparison between DNS and EDQNM results 
at moderate Reynolds number.
In the context of quasi-static MHD turbulence, it is hardly possible with 
DNS to reach high Reynolds number simulations without 
encountering artificial effects of the periodic boundary conditions, in view of 
the rapidly increasing numerical cost of 
pseudo-spectral methods with $\textit{Re}$.
%As already mentioned, one has to be careful when performing homogeneous tri-periodic DNS in an anisotropic context.
%Large correlation lengths of the velocity field can result in spurious confinement effects due to the periodicity.
In this section, we shall compare statistics obtained from the 
flow field predicted by DNS 
with data directly derived from EDQNM models, for a 
Reynolds number attainable by DNS.
We first present a comparison of dynamical quantities in section~\ref{sec:energetics},
then an extended analysis of anisotropy in section section~\ref{sec:aniso}.
\subsection{Energetics}
%--------------------------------
\label{sec:energetics}
%x
Total kinetic energy and total enstrophy
are presented in figures \ref{fig:kinetic-energy}(a) and (b) respectively.
After initialisation, the EDQNM model instantaneously builds triple 
correlations, or, in other terms, energy transfer spectra,
close to the ones observed in DNS.
Therefore, the initial dynamics 
for the EDQNM1-2 models and DNS are similar.
At larger times $t^*>1$, and for $N=1$, EDQNM2 (\textit{resp.} EDQNM1) 
seems to overestimate (\textit{resp.} underestimate) the value of kinetic energy.
For $N=5$, the kinetic energy and enstrophy decays
predicted by EDQNM2 and DNS are in good agreement. 
In both cases, it appears that EDQNM1 underestimates the kinetic energy
and the enstrophy after the initial short time stage, whereas EDQNM2
predicts decay rates that are remarkably close to the DNS evolution,
if one considers all the possible sources of statistical inaccuracies
which may appear in DNS data. The good performance of EDQNM2 with respect
to EDQNM1 is clearly the sign that including the explicit effect
of anisotropic Joule dissipation in the nonlinear dynamics is crucial
for modelling quasi-static MHD turbulence.
The scale-dependent Joule dissipation timescale 
$\tau_M(k)=1/M_0$ can thus be compared to the turbulent
timescale $\tau(k)=\varepsilon^{-1/3}k^{-2/3}$, where
$\varepsilon$ is the kinetic energy dissipation. Equating
these timescales yields a given wavenumber $\k_M=M_0^3\varepsilon^{-1/2}$, say,
which separates Joule dissipation dominated scales $k<k_M$ from dominant
nonlinear dynamics $k>k_M$ ($1/k_M$ is the equivalent of the 
Ozmidov scale introduced in stably stratified turbulence). 
For our runs at $N=1$, $k_M=5$ initially,
and $k_M=50$ at the end of the simulation, whereas for the run $N=5$, 
the figures are $60$ at the beginning, and $800$ in the end.
This shows that, apart from the early stage of the $N=1$ case,
in all our simulations, the energetic scales are dominated by
ohmic dissipation (this is illustrated on figure~\ref{fig:angular-spectra}).

The axisymmetric EDQNM model is also valid for 
isotropic turbulence, but the numerical 
cost is considerably larger than that of
the classical fully isotropic model.
The results of  isotropic DNS (\textit{i.e} 
setting $B_0=0$), presented on figure~\ref{fig:kinetic-energy},  
are obtained from the same initial conditions, and
show that the decay of kinetic energy is faster in the QS MHD case than in  
isotropic turbulence due to the additional ohmic dissipation.
Concerning the evolution of enstrophy, one observes an initial increase 
for both EDQNM models in the isotropic case and in DNS, showing a
short-time re-adjustment which cannot occur when the interaction parameter
$N$ is large, since the magnetic effect catches up almost instantaneously.

\begin{figure}
\unitlength 0.5mm
\begin{picture}(300,110)
        \put(-6,-2){\includegraphics[height=100\unitlength]{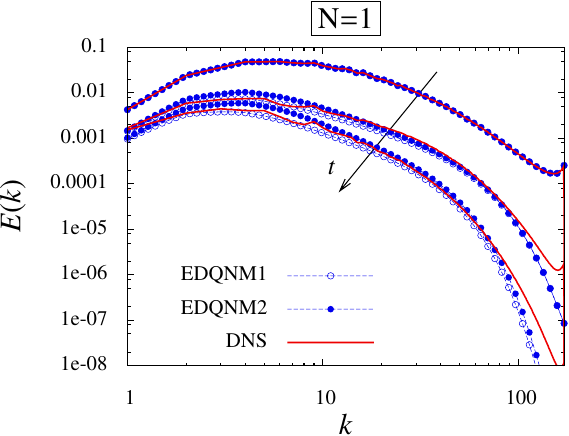}}
        \put(140,-2){\includegraphics[height=100\unitlength]{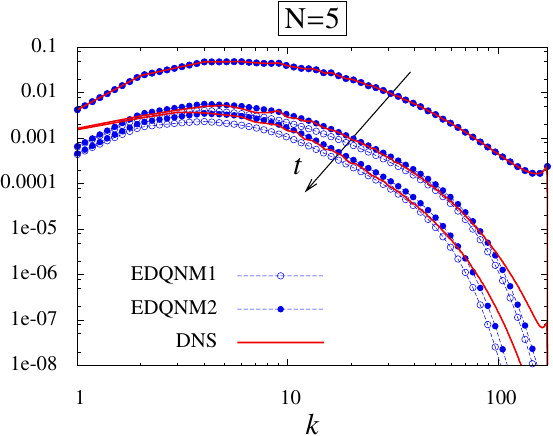}}
\end{picture}
\caption{\label{fig:energy-spectra} Kinetic energy spectra at dimensionless 
times $t^*=0$, $3$ and $6$.}
\end{figure}

The kinetic energy spectra are plotted in 
figure \ref{fig:energy-spectra} at 
three different times.
The initial energy spectra are identical, since EDQNM spectra are initialized 
from DNS results.
As already mentioned
the initial integral length scale (\textit{resp}. peak energy wave number) 
is smaller (\textit{resp.} larger) than for classical 
hydrodynamic simulations.
Figure \ref{fig:energy-spectra} shows that
the DNS and EDQNM2 spectral energy levels are 
in good agreement for all the scales of the flow.
For $N=1$ (figure~\ref{fig:energy-spectra}a), 
the slight overestimation of the energy by EDQNM2 is again 
observed, particularly at intermediate scales $6<k<12$, while
we retrieve the larger underprediction of the EDQNM1 model.
In all cases, the comparison between DNS and EDQNM
in the dissipative range of the spectrum is not as good 
independently of the value of $N$ and of the model version.
Several explanations can be put forward, both on the account
of the model or of the DNS approach: desaliasing in DNS, intermittency
not present in the EDQNM model, truncation in both, \textit{etc.}
%\begin{verbatim}
%Note however that the dealiasing process used here may be responsible for 
%an overestimation of the kinetic energy at small scales predicted by DNS.
%\end{verbatim}
%Finally, for $N=5$, the kinetic energy contained in small wave numbers is rather constant according to DNS results. 
Overall,  figure \ref{fig:energy-spectra} still
demonstrates that the EDQNM2 model is a good predictive model of the dynamics
of QS MHD over a wide range of scales.
\subsection{Refined comparison of the anisotropy}
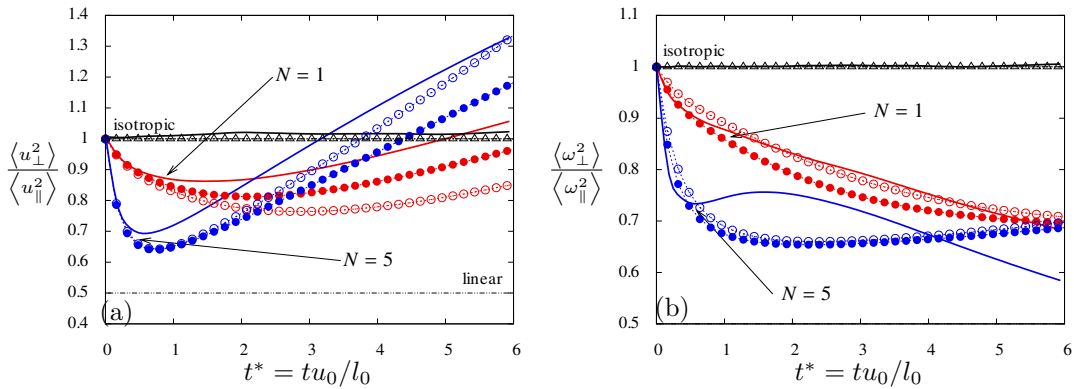
\begin{figure}
\unitlength 0.5mm
\begin{picture}(200,110)
        \put(-5,0){\includegraphics[height=100\unitlength]{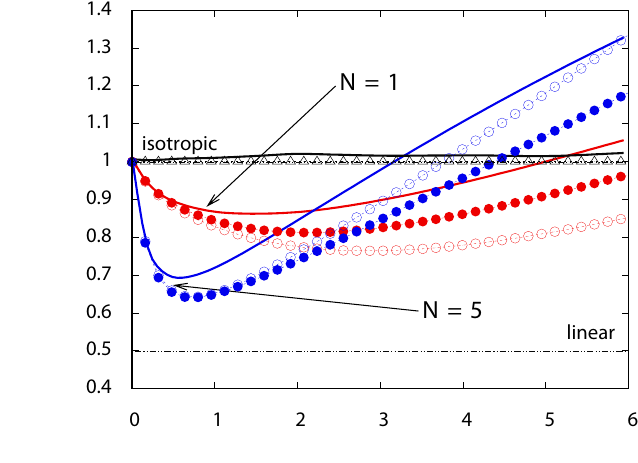}}
        \put(140,0){\includegraphics[height=100\unitlength]{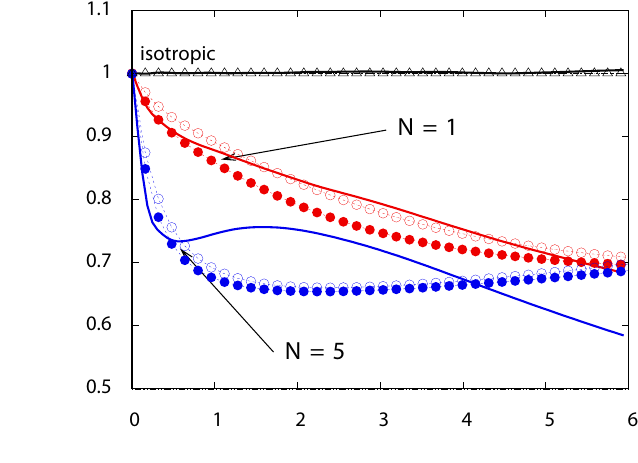}}
        \put(22,17){(a)}
        \put(168,18){(b)}
        \put(60,1){$t^*=tu_0/l_0$}
        \put(205,1){$t^*=tu_0/l_0$}
        \put(-3,55){$\frac{\left<u_\perp^2\right>}{\left<u_\parallel^2\right>}$}
        \put(140,55){$\frac{\left<\omega_\perp^2\right>}{\left<\omega_\parallel^2\right>}$}
\end{picture}
\caption{(a) Ratio between horizontal and vertical kinetic energies versus 
dimensionless time $t^*=tu_0/l_0$. (b) Ratio between horizontal and 
vertical enstrophies versus dimensionless time $t^*=tu_0/l_0$. --- DNS, 
$\circ$ EDQNM1, $\bullet$ EDQNM2, $\triangle$ Isotropic EDQNM.\label{fig:uvw}}
\end{figure}

%-----------------------------------------------------
\label{sec:aniso}
The level of anisotropy in the flow can be quantified with increasing
refinement degrees.
A first measure  is the ratio between horizontal 
and vertical kinetic energies, plotted in figure \ref{fig:uvw}(a).
The linear and inviscid regimes are characterised by the following 
scaling (\cite{moff67})
\vspace{-1mm}
\begin{equation}
\label{eq:moff_pred}
\left<u_\parallel^2\right>\simeq 2\left<u_\perp^2\right> ,
\end{equation}
where $u_\parallel=u_z$ is the axial velocity component, and $u_\perp=\sqrt{u_x^2+u_y²}$, with $u_x$ and $u_y$ 
the transverse
velocity components.
As already observed by \cite{voro05},  \cite{bura08_2}, \cite{favier10}, 
this linear prediction 
is not observed in numerical simulations.
The initial stage ($t^*<1$) is characterised by a decrease of the 
ratio $r_e = \left<u_\perp^2\right>/\left<u_\parallel^2\right>$, in agreement with 
equation~(\ref{eq:moff_pred}), but after a few turnover times, 
this ratio increases.
It was shown that this is not due to a restoration of isotropy but 
to a nonlinear phenomenon linked to the particular quasi-two-dimensional 
structure of the flow 
(for details, see \cite{favier10} and section \ref{sec:2D3C}).
Figure \ref{fig:uvw}(a) shows that EDQNM2 
reproduces this departure from the linear prediction, 
although with a time lag and a smaller amplitude.
At small interaction parameter, 
EDQNM2 provides a better agreement 
with DNS than EDQNM1 for $N=1$, less so for $N=5$.
%but this is only due to the addition of diverging sources of departure
%at this specific value of $N$.

The ratio between transverse and axial kinetic energies presented on 
figure~\ref{fig:uvw}(a) sets the focus on the large scale dynamics.
The small scale dynamics can be brought forward 
by computing  a similar quantity based on vorticity components. 
We define the ratio between 
transverse and axial enstrophies as
\begin{equation}
\label{eq:ratio_enstr}
r_\omega=\frac{\left<\omega_\perp^2\right>}{\left<\omega_\parallel^2\right>} \ .
\end{equation}
%x
In a pure two-dimensional case, 
this ratio goes to zero, whereas in the isotropic case, 
it is about one.
For $N=1$ on figure~\ref{fig:uvw}(b), 
the ratio is always decreasing independently of the model considered, 
but is far from the two-dimensional value.
For $N=5$, there is a clear departure between DNS and EDQNM predictions. 
Initially in DNS, there is a strong decay of $r_\omega$, then the trend is reversed
synchronously with the decay reversal of $r_e$ (figure \ref{fig:uvw}a), 
at $t^*\approx 1$. Eventually, $r_\omega$
decreases again.
This three-stage evolution is not captured by the EDQNM model. 
The first increase stage after the initial decrease is reproduced,
with a delay as for $r_e$, but the second
change of slope is not.
It seems that a phenomenon appears in DNS at $t^*\approx2-3$,
whereby  the ratio $r_\omega$  decreases in DNS,
which is not captured by the model. The multiplicity of possible
nonlinear time scales in MHD turbulence might not be reproduced
by the single time scale introduced in the closure (equation~\ref{mu}).

Let us now compare the DNS and EDQNM results concerning the prediction of 
the directional anisotropy resulting from  ohmic dissipation.
The first effect of the magnetic field is to 
dissipate preferentially Fourier modes 
with wave vector $\bm{k}$ parallel to $\bm{B}_0$.
A direct consequence is the decrease of the transverse
kinetic energy with respect to the 
axial one (since modes with $\bm{k}\parallel\bm{B}_0$ contribute only 
to transverse energy, 
see figure \ref{fig:craya}).
This is observed in figure \ref{fig:uvw}(a).

The simplest way to quantify this 
directional anisotropy (\textit{directivity}) 
is to consider
%
%
%
%
%
%
%***** 
%
typical angles defined in physical space, such as the ones introduced 
by Moreau and Shebalin (\cite{ALEMANY-MOREAU-SULEM-FRISCH,sheb83}).
The `Moreau angle' $\beta$ defined by
\begin{equation}
\cos^2 \beta (t)= ({\cal K}(t))^{-1} \iiint \cos^2 \theta e(\vk,  t) \mathrm{d}^3 \vk,
\label{eq:moreau}
\end{equation}
directly derives from the  one-point dynamical equation for the kinetic energy 
${\cal K}(t) = \iiint e(\vk, t)\mathrm{d}^3\vk$
\begin{equation} 
{d}{\cal K}/dt + 2M^2_0 \cos^2 \beta {\cal K} = - \varepsilon
\label{eq:tke} \end{equation}
coming from integration of~equation~(\ref{line}). This equation suggests
as well to refine the definition of
the separating
wavenumber introduced in section~\ref{sec:energetics}
as $k_M=M_0^3\cos^3\beta\epsilon^{-1/2}$.

\begin{figure}
\unitlength 0.5mm
\begin{picture}(300,110)
        \put(-2,-2){\includegraphics[height=100\unitlength]{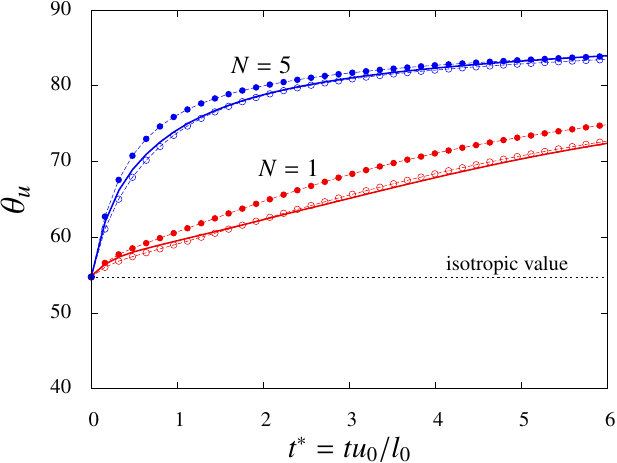}}
        \put(140,-2){\includegraphics[height=100\unitlength]{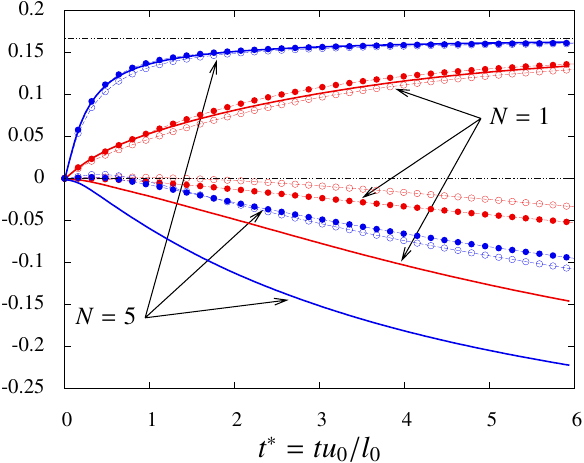}}
        \put(268,85){\makebox(0,0){$\left.\rule[-2mm]{0mm}{4mm}\right\}$}}
        \put(268,33){\makebox(0,0){$\left.\rule[-4mm]{0mm}{16mm}\right\}$}}
	\textcolor{black}{\put(271,84){\scriptsize{$b_{33}^e$}}}
	\textcolor{black}{\put(271,30){\scriptsize{$b_{33}^{\mathcal{Z}}$}}}
        \put(15,18){(a)}
        \put(154,18){(b)}
\end{picture}
\caption{\label{fig:b33} Anisotropic coefficients versus dimensionless time 
$t^*=tu_0/l_0$. (a) Shebalin angle $\theta_u$. (b) Anisotropic part of the 
Reynolds stress tensor $b_{33}=b_{33}^e+b_{33}^{\mathcal{Z}}$. --- DNS, 
$\circ$ EDQNM1 and $\bullet$ EDQNM2.}
\end{figure}
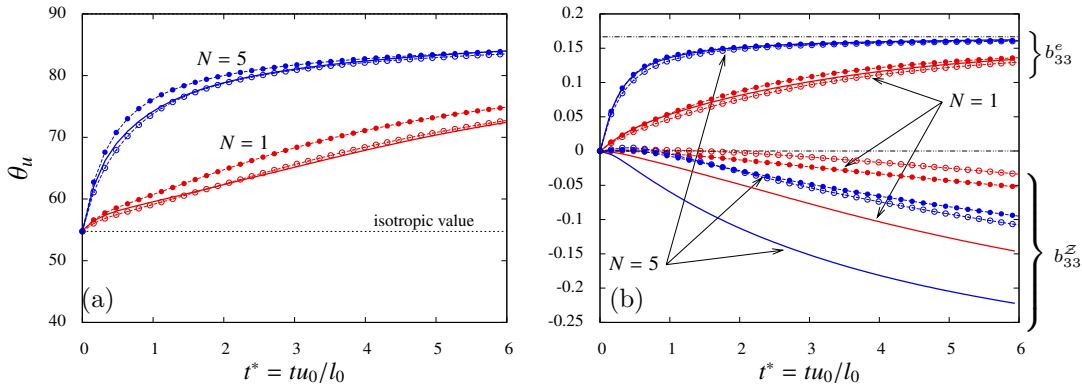

The Shebalin angle, more widely used in the MHD community, 
 characterizes
the angular distribution of the vorticity spectrum $k^2 e$, as
evidenced by its definition contrasted with equation~(\ref{eq:moreau}):
\begin{equation}
\cos^2 \theta_u (t) = (\langle \omega^2 \rangle(t))^{-1} \iiint k^2 \cos^2
\theta e(\vk, t) \mathrm{d}^3 \vk,
\label{eq:sheb2} 
\end{equation}
where the enstrophy is  $\langle \omega^2 \rangle = \iiint k^2 e(\vk, t) \mathrm{d}^3 \vk$.
This definition is the continuous counterpart, in a slightly 
different form,  of the classical discretized
version \cite{sheb83}, used  for the plots in figure \ref{fig:b33}(a):
\begin{equation}
\tan^2\theta_u=\frac{\sum_{\bm{k}}k_\perp^2|\hat{\bm{u}}(\bm{k},t)|^2}{\sum_{\bm{k}}k_\parallel^2|
        \hat{\bm{u}}(\bm{k},t)|^2} \ ,
\label{eq:sheb}
\end{equation}
where 
$k_\perp=\sqrt{k_x^2+k_y^2}$ is the transverse component of the wave vector, 
and $k_\parallel=k_z =k^2 \cos^2 \theta$ is the axial one. On the other hand, we do not plot
directly the Moreau angle here, but the equivalent quantity
$b^e_{33}$ defined by equation~(\ref{eq:b33e}): $b^e_{33}$ is proportional 
to the intensity of the first angular harmonic of $e$,
through
%
%\begin{equation}
$b^e_{33} = {1}/{6} - (1/2)\cos^2 \beta.$
%\label{eq:betobeta} 
%\end{equation}
%

The Shebalin angles for the velocity field are first plotted 
in figure \ref{fig:b33}(a).
In all cases ---and similarly for the Moreau angles---, the increase from the isotropic initial value
 $\theta_u\approx 54.7^{\circ}$  
indicates a 
concentration of energy in modes perpendicular to the 
imposed magnetic field. This two-dimensional limit corresponds to
$\theta_u\approx 90^{\circ}$.
This is a well-known consequence of the ohmic dissipation, which results in 
physical space in a flow invariant in the axial direction.
We note that EDQNM2 overpredicts 
the value of the Shebalin angle with respect to EDQNM1 and DNS. This overestimation does not concern
$b^e_{33}$, as shown in
figure \ref{fig:b33}(b). This suggests that the EDQNM prediction for
the directional anisotropy is different for larger scales (energy
distribution) and smaller
scales (vorticity distribution), with a particular sensitivity of EDQNM2
at smaller scales. A small inaccuracy can therefore pull the Shebalin 
angle predicted by EDQNM2 in the wrong way, even if EDQNM2 gives a better
overall prediction than EDQNM1. 

The \textit{polarization}   anisotropy is another kind
of anisotropy that may appear in addition to the directional anisotropy.
This anisotropy cannot be quantified with Shebalin angles, since it is
not directly related to the dependence of the poloidal and toroidal
velocity components upon $\theta$, but is related to their difference.
Its characterization requires a specific splitting of the 
deviatoric part $b_{ij}=R_{ij}/(2\mathcal{K})-\delta_{ij}/3$ 
of the Reynolds tensor  
$R_{ij}=\left<u_i(\bm{x})u_j(\bm{x})\right>$,
where
$\mathcal{K}$ is the total kinetic and 
$\delta_{ij}$  the Kronecker tensor.
Considering the axisymmetry of the flow about the axis of $\bm{B}_0$, 
only one diagonal term is needed to describe the anisotropy, $b_{33}$ say.
Using equations~(\ref{eq:poltor}) and~(\ref{eq:eZH}), one obtains the
two contributions for $b_{33}=b_{33}^e+b_{33}^{\mathcal{Z}}$ 
(\cite{camb89,CAMBON-MANSOUR-GODEFERD}), with:
%\vspace{-2mm}
\begin{align}
b_{33}^e & =\frac{1}{2\mathcal{K}}
     \int\left(e(\bm{k})-\frac{E(k)}{4\pi k^2}\right)
     \sin^2\theta\textrm{d}^3\bm{k} \label{eq:b33e} \\
\label{eq:b33z}
b_{33}^{\mathcal{Z}} & =\frac{1}{2\mathcal{K}}\int\mathcal{Z}(\bm{k})
    \sin^2\theta\textrm{d}^3\bm{k}
\end{align}
where $\theta$ is the polar angle between the wave vector $\bm{k}$ and the 
axis of symmetry (see figure~\ref{fig:craya}), $E(k)$ is the 
spherically-averaged kinetic energy spectrum, $\mathcal{Z}(k)$ 
is the polarization spectrum.
As stated by its definition~(\ref{eq:b33e}), 
$b_{33}^e$ is similar to the Shebalin angles in that it quantifies 
the directivity of the energy with respect to the vertical direction. 
$b_{33}^{\mathcal{Z}}$ quantifies the additional dimensionality anisotropy
which is conveyed by the polarization spectrum $\mathcal{Z}$.
The limiting value $b^e_{33} = 1/6$ is reached for 
two-dimensional flows, in both the 2D-3C and the 2D-2C cases,
distinguished only by the value of $b^{Z}$: $0$
for 2D-3C flows, $-1/2$ for 2D-2C flows.

Figure \ref{fig:b33}(b) presents 
the evolution of $b_{33}^e$ and $b_{33}^{\mathcal{Z}}$ versus time.
Concerning $b_{33}^e$, the same conclusions as 
the ones resulting from  the Shebalin angles are drawn from the figure.
Note that it is possible to rescale time with the 
ohmic dissipation characteristic time so that both $b_{33}^e$ and $\theta_u$ 
collapse independently of the intensity of $B_0$ (\cite{favier10}
and analytical law in appendix~\ref{appendixB}).

The polarization part $b_{33}^{\mathcal{Z}}$, 
which is zero initially, decays in all cases, showing a global
predominance of toroidal over poloidal energy.
Negative polarization is thus responsible for the increase of the 
componental enstrophy and velocity ratios $r_\omega$ and $r_e$
plotted in figure \ref{fig:uvw}.
EDQNM1-2 models underestimate the amplitude of polarization, 
%but we
%point out that it is a feature rather difficult to predict,
%since it quantifies the flow anisotropy related to its
%dimensionality, \textit{i.e.}, in short, the predominance of gradients in one
%or more directions of space.
but this is not necessarily a defect of the closure, 
given the spurious confinement effects yielding polarization
in DNS, as shown and discussed in section~\ref{sec:conf}. 
As already observed, EDQNM2 is in better agreement with DNS for $N=1$ 
(when nonlinearities are important) whereas EDQNM1 compares better for $N=5$ 
(when nonlinearities are dominated by ohmic dissipation).
In view of the value of the separation
scale $k_M$ presented in section~\ref{sec:energetics}, 
the dynamics is driven by nonlinear timescale only at the
beginning of the simulation at $N=1$. As mentioned in the Introduction,
the additional physics injected into the EDQNM2 model through the  straining
timescale (see Appendix~\ref{appendixA}) is corrected by the
Joule dissipation time-scale, but the imbalance of the two acts
variably depending on the regime. It seems here that the EDQNM2
nonlinear improvements are too large for these Joule dissipation
dominated scales.

% Note also that, as previously discussed in section~\ref{sec:conf}, 
% confinement due to 
% periodic boundary conditions in DNS is directly responsible for a decrease of 
% the polarization and might be responsible for some of the differences 
% observed in figure \ref{fig:b33}(b). 

%
\begin{figure}
\unitlength 0.5mm
\begin{picture}(300,110)
        \put(-2,-2){\includegraphics[height=100\unitlength]{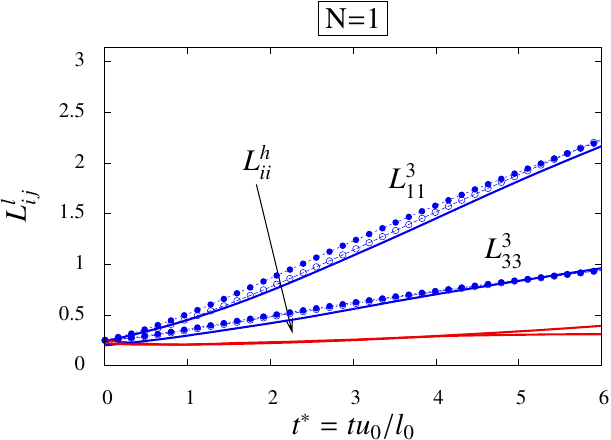}}
        \put(140,-2){\includegraphics[height=100\unitlength]{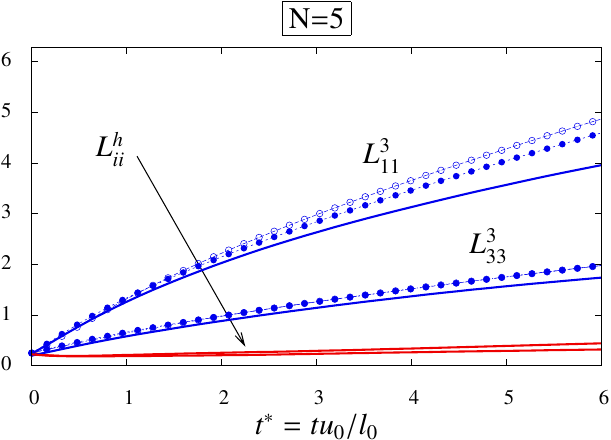}}
\end{picture}
\caption{\label{fig:lij} Velocity correlation lengths for 
$N=1$ (left) and $N=5$ (right). Superscripts $3$, $h$ and $i$ 
correspond to vertical direction (aligned with $\bm{B}_0$), 
horizontal direction and any direction, respectively. --- DNS, 
$\circ$ EDQNM1 and $\bullet$ EDQNM2.}
\end{figure}

\begin{figure}
\unitlength 0.5mm
\begin{picture}(300,100)
        \put(50,-2){\includegraphics[height=100\unitlength]{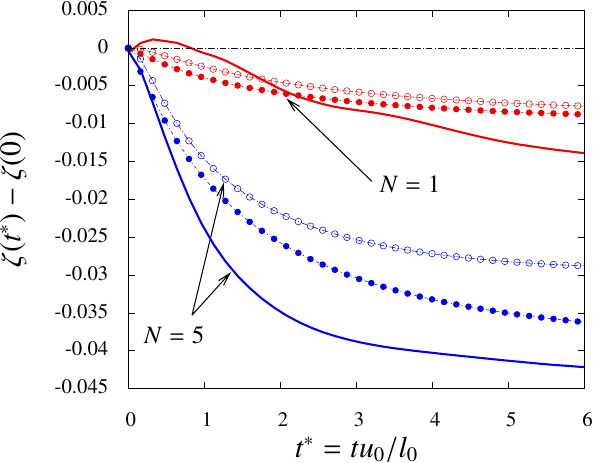}}
\end{picture}
\caption{\label{fig:uilij} Evolution with time of 
$\zeta=\langle u_3^2\rangle L_{33}^{(3)}-2\langle u_1^2\rangle L_{11}^{(3)}$. --- DNS, $\circ$ EDQNM1 and $\bullet$ EDQNM2.}
\end{figure}

We then consider the time evolution of the velocity correlation 
lengths defined by equations~(\ref{eq:deflijl}), presented
on figure \ref{fig:lij}.
At the end of the simulations ($t^*\approx6$), 
$L_{33}^{(3)}\approx2.2$ and $0.98$ 
for the respective cases $N=5$ and $1$.
The axial correlation length of axial velocity
$L_{33}^{(3)}$ is therefore always 
significantly smaller than the box size $2\pi$.
The axial correlation length of transverse velocity, however,
for the case at $N=5$,  
reaches about two thirds of the 
numerical box size.
The correlation lengths obtained from EDQNM are close to the ones 
computed from DNS results, indicating a good prediction of the anisotropy
of the large scale structures of the flow.
As previously discussed in section \ref{sec:conf}, 
the growth of $L_{11}^{(3)}$ 
computed by DNS seems to slow down in time, 
a fact that can be attributed to the periodic 
boundary conditions. Such saturation is not apparent in EDQNM results, so 
that the correlation lengths continue to grow. The confinement-related
explanation is supported by the similitude 
between figure \ref{fig:lijlin}(b) and \ref{fig:lij}(b).

Moreover, it is possible to isolate the contribution due to polarization 
by computing
\begin{equation}
\label{eq:2dcomp}
\zeta=\langle u_3^2\rangle L_{33}^{(3)}-2\langle u_1^2\rangle L_{11}^{(3)}
   =\int_0^{\infty}4\pi^2\Re\mathcal{Z}(\bm{k})\Big|_{k_z=0}k\textrm{d}k \ .
\end{equation}
%
%plotted on figure \ref{fig:uilij}.
This quantity is interesting for two reasons: (a) from equations \eqref{eq:defl333} 
and \eqref{eq:defl113}, its departure from zero is only due to the 
polarization $\mathcal{Z}(\bm{k})$; (b) this quantity is accessible 
experimentally.
Initially, $\zeta$ is exactly zero for EDQNM models since the 
polarization is set to zero at the beginning of the calculation.
However,  $\zeta(t^*=0)=-0.015$ in DNS is small
but not exactly zero.
This may be a trace of the forcing scheme used to reach a quasi-steady state of 
hydrodynamic turbulence for $t^*<0$.
$\zeta$ may also be dominated by 
contributions from small values of $k_z$, where the DNS 
spectral discretization 
is too coarse to yield converged statistics.
In all cases, figure~\ref{fig:uilij} shows that $\zeta(t^*)-\zeta(0)$ decreases, 
%\marginpar{\textcolor{red}{figure 9: replot $\zeta(t^*)-\zeta(0)$}}
in accordance with 
negative polarization.
One observes that the equatorial polarization is underpredicted by EDQNM, 
which is consistent with the previous observations 
on the deviatoric tensor $b_{33}$. 
However, the relative evolutions of the $N=1$ and $N=5$ EDQNM
predictions for $\zeta$ agree correctly with the dependence with $N$
observed on the DNS curves.

\begin{figure}
\unitlength 0.5mm
\begin{picture}(200,110)
        \put(0,-5){\includegraphics[height=100\unitlength]{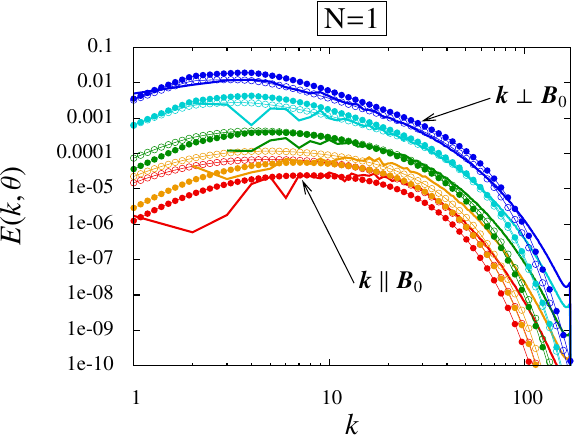}}
        \put(140,-5){\includegraphics[height=100\unitlength]{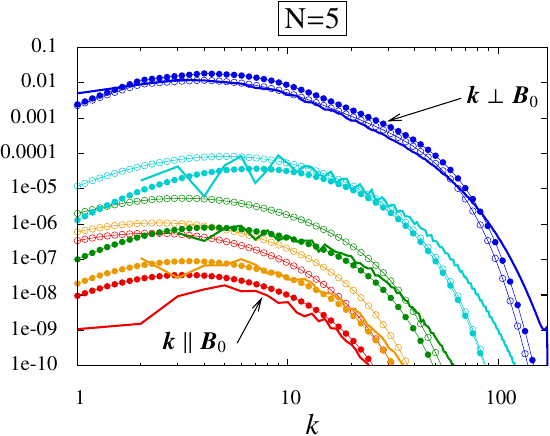}}
\end{picture}
\caption{\label{fig:angular-spectra} Angular energy spectra at $t^*=5$. --- DNS, 
$\circ$ EDQNM1 and $\bullet$ EDQNM2. In each case, five curves are plotted, 
from top/equator ($\bm{k}\perp\bm{B}_0$) to bottom/pole 
($\bm{k}\parallel\bm{B}_0$). $N=1$ (left) and $N=5$ (right).}
\end{figure}

All the previous statistics involve a spectral integration over wave numbers, 
so that information about scale dependency is lost.
On the contrary, the angular 
spectrum $E(k,\theta)$
retains both scale- and angle-dependence:
\begin{equation}
\label{eq:ekt_def2}
E(k,\theta)=\left[
     \int_{\theta-\Delta\theta/2}^{\theta+\Delta\theta/2}\cos\theta\mathrm{d}\theta\right]^{-1}
     \sum_{\substack{k-\Delta k/2<|\bm{k}|<k+\Delta k/2 \\ 
     \theta-\Delta\theta/2<\theta<\theta+\Delta\theta/2}} 
     \hat{u}_i(k,\theta)\hat{u}^*_i(k,\theta) \ .
\end{equation}
where $\Delta k$ and $\Delta\theta$ specify the discretization steps in 
Fourier space used for computing the anisotropic spectra 
(see figure \ref{fig:craya} in which
the shaded region corresponds to the scales which 
contribute to $E(k,\theta)$).
Ring-averaged angular spectra $E(k,\theta)$ have 
already been used in the context 
of rotating turbulence by \cite{CAMBON-MANSOUR-GODEFERD} and 
for stably stratified turbulence by \cite{GODEFERD-STAQUET}, 
and are similar to the 
ring decomposition by \cite{bura08}.
We choose here $\Delta k=1$ and $\Delta\theta=\pi/10$,
figures that depend on the DNS resolution to ensure optimal 
statistical sampling.
%In EDQNM closures, the polar angle is discretized in a different way so that 
%one has to interpolate the values for comparison with DNS statistics.
The angular spectra are plotted on figure \ref{fig:angular-spectra}, 
at time $t^*=5$.
At the initial time $t^*=0$, 
all angular spectra collapse since the initial condition is isotropic.
Figure~\ref{fig:angular-spectra} shows that, 
as time increases, most of the kinetic energy is concentrated 
in the spectrum with transverse wavevectors, 
since the Joule dissipation term in equation~(\ref{line})
reduces less energy at this orientation,
independently on the wavenumber.
The qualitative agreement of EDQNM model predictions
with the DNS ones is impressive, considering the multi-scale, multi-directional
character of these spectral statistics.
There are, however, some differences.
First, one observes that EDQNM2 overestimates slightly
 the equatorial kinetic energy, 
which is consistent with the overestimation of the Shebalin angle already 
observed in figure \ref{fig:b33}(a).
However, the global angular dependency of the energy observed in DNS is well 
reproduced by EDQNM2, whereas EDQNM1 overestimates the polar kinetic energy 
(see lowermost curves with $\circ$ symbols on figure \ref{fig:angular-spectra}).
In all models, as $N$ increases, the angular anisotropy increases so that the 
flow tends to be invariant in the vertical direction.

\begin{figure}
\unitlength 0.5mm
\begin{picture}(300,100)
        \put(-10,-5){\includegraphics[height=100\unitlength]{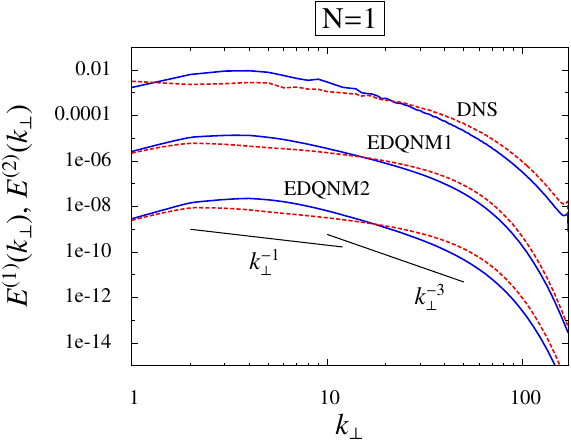}}
        \put(140,-5){\includegraphics[height=100\unitlength]{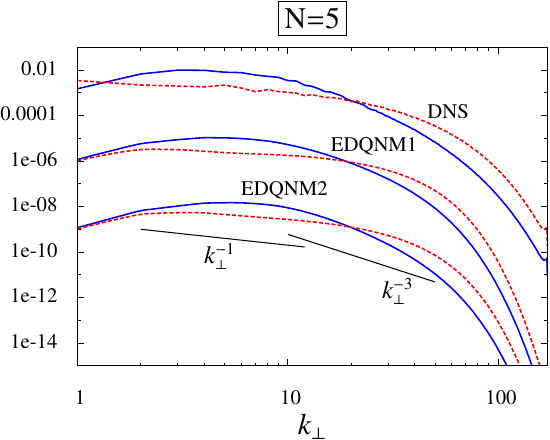}}
\end{picture}
\caption{\label{fig:spange}Equatorial energy spectra at $t^*=6$. 
The solid blue lines correspond to $\Phi^{(1)}(k_{\perp})$,
the dotted red lines 
correspond to $\Phi^{(2)}(k_{\perp})$. EDQNM1 results are shifted down by 
three decades, EDQNM2 ones are shifted down by six decades.}
\end{figure}

So far, we focused on the angular dependency of the kinetic energy.
The departure from isotropically distributed energy is due to Joule dissipation and 
is observable in the growth of the Shebalin angle $\theta_u$ 
(see figure \ref{fig:b33}(a)), $b_{33}^e$ (see figure \ref{fig:b33}(b)), 
and in angular spectra.
However, it has been demonstrated that this effect is mostly linear,
 and that it can explain neither
 negative values of $b_{33}^{\mathcal{Z}}$ (see figure \ref{fig:b33}(b)), nor  
the increase of the ratio between transverse and axial energies at large times 
(see figure \ref{fig:uvw}).
The poloidal/toroidal decomposition of spectral quantities (equation~\eqref{eq:poltor}), along
with the angular spectral distribution, provides a way of understanding
these unexplained features.
Figure \ref{fig:spange} presents the equatorial spectra (\textit{i.e.} only transverse 
wave vectors are considered) decomposed as poloidal (\textit{i.e.} axial in this 
particular case, see figure \ref{fig:craya}) and toroidal (\textit{i.e.} transverse
in this configuration) 
contributions.
%The DNS results are on the top of the figure, the EDQNM1 and EDQNM2 results are 
%shifted down by 3 and 6 decades respectively for clarity.
The \textit{polarization} anisotropy is clearly observable, as the difference
between the two spectra.
It is scale-dependent, with negative polarization at large scales 
($\Phi^{1}>\Phi^{2}$), 
responsible for the negative value of $b_{33}^{\mathcal{Z}}$ 
%(since $b_{33}^{\mathcal{Z}}$ is the summation of the difference between 
%poloidal and 
%toroidal energies over all spectral space); 
and  positive 
polarization at small scales 
($\Phi^{1}<\Phi^{2}$).
The structure of the flow is therefore strongly scale-dependent with  
dominance of transverse kinetic energy at large scales and a dominance 
of axial kinetic energy at small scales.
This departure from the poloidal/toroidal equipartition of energy is 
mainly observable for transverse wavevectors, 
where the energy accumulates because of ohmic dissipation.
For axial wavevectors,  $\theta$ goes to zero and this is no longer observable.
Note that the cross-over wave number $k_\perp^c$ at which 
$\Phi^{1}(k_{\perp}^c)= \Phi^{2}(k_{\perp}^c)$ 
($k_{\perp}^c\approx \textcolor{black}{20}$ on figure \ref{fig:spange}) depends mainly on 
the initial conditions and on the Reynolds number.
Both EDQNM1-2 models reproduce this 
non-linear behaviour as well as the approximate location
of the cross-over  wave number.
%

%\begin{figure}
%\unitlength 0.5mm
%\begin{picture}(300,100)
%%
%        \put(-10,-5){\includegraphics[height=100\unitlength]{./spz1}}
%        \put(140,-5){\includegraphics[height=100\unitlength]{./spz5}}
%%
%\end{picture}
%\caption{\label{fig:spz}Difference between poloidal and 
%toroidal energy spectra at $t^*=6$. --- DNS, $\circ$ EDQNM1 and $\bullet$ EDQNM2.}
%\end{figure}
%%

$k_\perp^{-3}$ and $k_\perp^{-1}$ slopes are indicated on figure~\ref{fig:spange}
for comparison with common scalings of 
two-dimensional turbulence with passive scalar 
(see \cite{batch59}, \cite{bos09}, and the discussion of 
the analogy with two-dimensional three components flows in section \ref{sec:2D3C}).

%Note that the previous results concerning the poloidal/toroidal decomposition 
%do not explain why the polarization (which is basically due to difference between 
%poloidal and toroidal energies) is underpredicted by EDQNM closures 
%(see figure \ref{fig:b33}(b)).
%$b_{33}^Z$ is simply derived by integration over all wave numbers of the 
%difference $E^{(2)}(k)-E^{(1)}(k)$.
%In the quasi-static case, the dominant contribution of this quantity is from 
%equatorial modes because of ohmic dissipation.
%In figures \ref{fig:spz}(a) and \ref{fig:spz}(b), we present the difference 
%$E^{(2)}(k)-E^{(1)}(k)$ at $t^*=5$ for the cases $N=1$ and $N=5$.
%One observes that the dominant contributions to the polarization come from the 
%range $k<10$.
%In all cases, the polarization obtained from EDQNM closures is less important 
%than the one observed in DNS, which might be due to confinement.

\subsection{Dynamical equilibrium and energy transfer spectra}
\begin{figure}
\unitlength 1mm
\begin{picture}(130,42)
\put(0,0){\includegraphics[height=40mm]{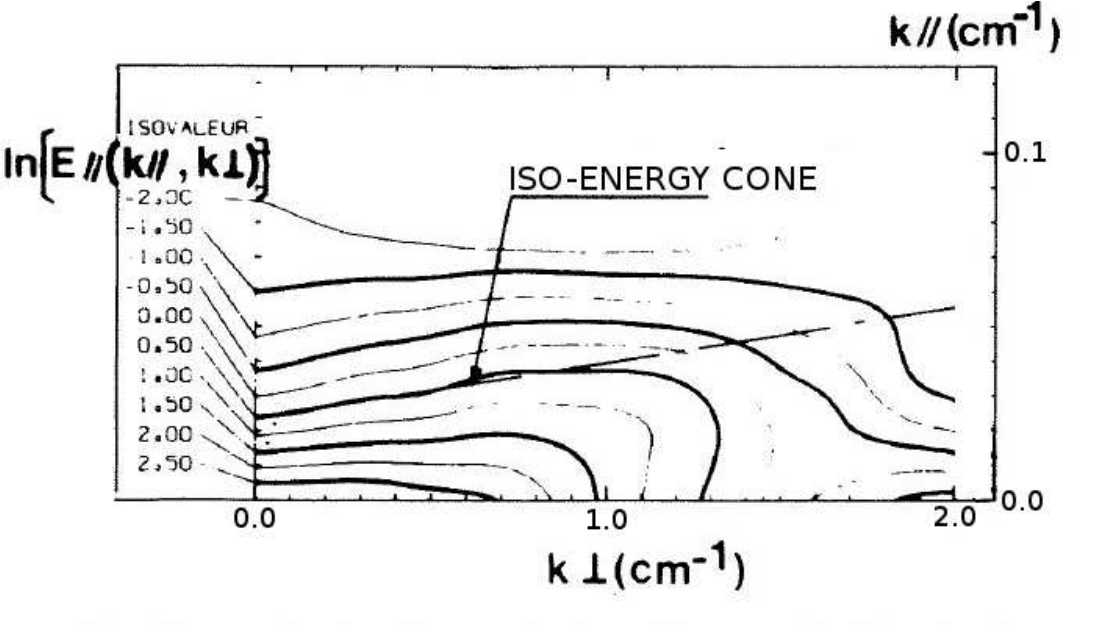}}
\put(70,6){\includegraphics[height=31mm]{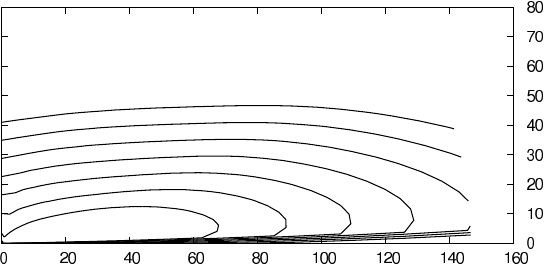}}
\put(100,2){$k_\perp$}
\put(135,20){$k_\parallel$}
\put(55,33){(a)}
\put(115,33){(b)}
\end{picture}
\caption{\label{fig:caperan} Iso-contours of the two-dimensional spectral distribution
of the axial kinetic energy $E_\parallel(k_\perp,k_\parallel)$ in logarithm scale.
(a) Figure extracted from \cite{CAPERAN-ALEMANY}, at
$\textit{Re}\simeq 1800$ and the interaction parameter $N\simeq 0.6$. The conical
spectral distribution is shown. (b) EDQNM2 
result at $t^*= 12$, $\textrm{Re}=333$ and $N=5$,
%\textcolor{red}{Benjamin: j'ai utilis\'e le champ 41 du Confin2/3 sur ecu...}
with 
iso-contour levels from -60 to -20 by steps of 5.}
\end{figure}
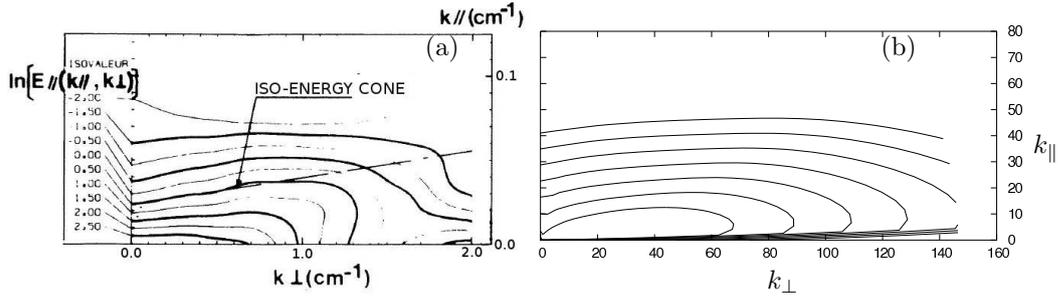

\begin{figure}
\unitlength 0.5mm
\begin{picture}(300,110)
        \put(-6,-2){\includegraphics[height=100\unitlength]{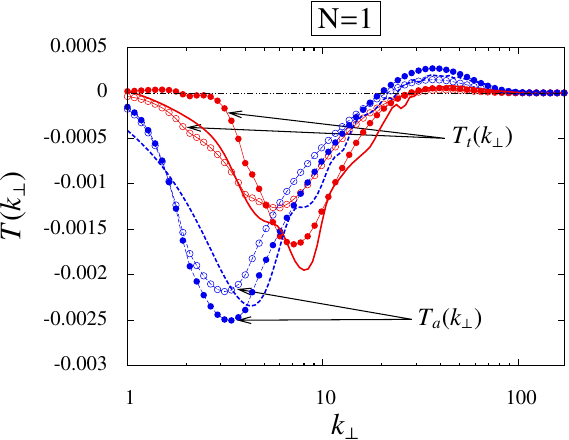}}
        \put(140,-2){\includegraphics[height=100\unitlength]{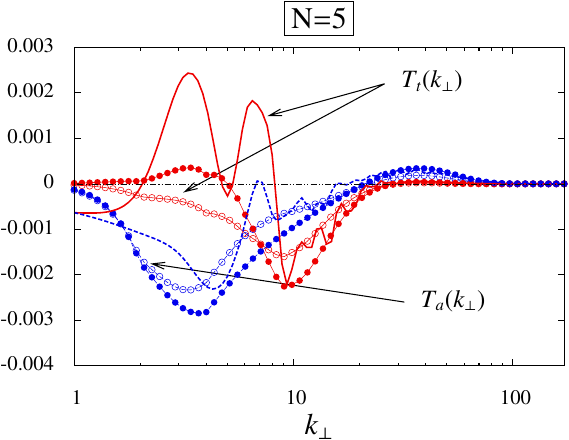}}
\end{picture}
\caption{\label{fig:energy-transfer} Equatorial kinetic energy transfer 
spectra at dimensionless times $t^*=5$. --- DNS, $\circ$ EDQNM1 and $\bullet$ EDQNM2.}
\end{figure}

%--------------------------------------
%
The anisotropic re-distribution of energy in quasi-static MHD
turbulence, starting from isotropic
initial turbulence, is the result of an essentially angular transfer,
as we have shown above with DNS and the EDQNM model, and as was observed
in towed-grid turbulence in mercury by \cite{ALEMANY-MOREAU-SULEM-FRISCH,CAPERAN-ALEMANY}.
These authors, using interaction parameters between $N\simeq 0.6$ and $1.17$, also observe
the appearance of a $k_\parallel^{-3}$ scaling for the axial kinetic energy spectrum
$E_\parallel(k_\parallel)$, that progressively replaces the Kolmogorov scaling $k_\parallel^{-5/3}$
over an increasingly wider wavenumber range. 
The complete $E_\parallel(k_\perp,k_\parallel)$ distribution, plotted on figure~\ref{fig:caperan},
pictures the spectral equilibrium of energy, due to both Joule dissipation---that 
drains energy towards the transverse 2D plane---and nonlinear inertial transfers.
As argued by~\cite{CAPERAN-ALEMANY}, the equilibrium
between the two phenomena should lead to a conical distribution of spectral energy,
which seems to be observed on figure~\ref{fig:caperan}(a).
The same quantity computed with EDQNM is plotted on figure~\ref{fig:caperan}(b).
The model permits this refined representation since it provides a smooth
distribution of the spectra, hardly available in DNS. The comparison
between the two panels of figure~\ref{fig:caperan} suggests strong similarities
in the dynamical equilibrium obtained in the experiment and in the EDQNM model.
[From figure~\ref{fig:angular-spectra} which presents angular spectra, but
contains the same information as shown differently on figure~\ref{fig:caperan}(b), 
we believe that an equivalent agreement would be obtained with DNS.]
One must bear in mind, however, that the dimensional 
scalings of both plots of figure~\ref{fig:caperan}
are different, so that no quantitative agreement is claimed.
In order to investigate further inertial transfers in the QS MHD turbulent
flow, we compute energy transfer spectra.
They are presented on figure \ref{fig:energy-transfer} 
at the same time $t^*=5$ as the spectra of figure~\ref{fig:angular-spectra}. 
In DNS, the spherically averaged transfer spectrum 
is directly computed from the nonlinear term $\bm{s}=\bm{u}\times\bm{\omega}$, 
with $\omega=\mathbf{\nabla}\times\vu$, as
\begin{equation}
T_i(k)=\sum_{k-\Delta k \le |\bm{k}| < k+\Delta k }\frac{1}{2}
  \left[\hat{u}_i(\bm{k})\hat{t}_i(-\bm{k})+\hat{u}_i(-\bm{k})\hat{t}_i(\bm{k})\right]
\end{equation}
where $\hat{\bm{t}}=-k^2\left[\bm{k}\times\left(\bm{k}\times\hat{\bm{s}}\right)\right]$. 
We focus here on equatorial modes $\bm{k}\perp\bm{B}_0$ and we distinguish the 
axial equatorial transfer $T_a(k_{\perp})$ and the transverse equatorial transfer 
$T_t(k_{\perp})$. In EDQNM closures, these quantities are directly obtained as
\begin{align}
T_t(k_{\perp})&=T^{(e)}(k,\theta=\pi/2)-T^{(Z)}(k,\theta=\pi/2)\\
T_a(k_{\perp})&=T^{(e)}(k,\theta=\pi/2)+T^{(Z)}(k,\theta=\pi/2) \ .
\end{align}
We observe an overall good agreement between DNS and EDQNM on 
figure~\ref{fig:energy-transfer}.
For $N=1$, one observes a reduced transverse transfer compared to the axial one, 
both in DNS and in EDQNM closures.
For $N=5$, DNS and EDQNM2 clearly display a positive transfer at large scales, 
characteristic of an inverse cascade of kinetic energy.
As described in \cite{favier10}, the transverse component of the velocity behaves 
as in two-dimensional turbulence, 
with the axial velocity component acting as a passive 
scalar, thus characterised by a classical direct cascade.
This inverse cascade of transverse velocity explains the reduction of dissipation 
and thus the dominance of transverse kinetic energy at large times 
(see figure \ref{fig:uvw}).
Note that EDQNM1 is unable to reproduce the inverse cascade observed in DNS and EDQNM2.
Finally, the oscillations observed in the DNS transfers for $N=5$ could be
explained by 
% an intermittency of the turbulent structures
% which is not present in the $N=1$ case, a phenomenon which is not included in 
% statistical models.
the fact that DNS yields one particular realization of the flow. The
statistics of a flow can differ significantly from what is
computed from an instantaneous flow field, in particular in the large scales.
We therefore do not exclude that the double positive lobe of $T_t(k_\perp)$
vanishes if we average over more flow realizations.

%
%%%%%%%%%%%%%%%%%%%%%%%%%%%%%%%%%%%%%%%%%%%%%%%%%%%%%%%%%%%%%%%%%%%%%%%%%%%%%%%%%%%%%%%
%
\section{Additional results accessible only with the EDQNM closure model}
%-----------------------------------------------------
\label{sec:edqnm}
We have presented in section~\ref{sec:comp} a comparison
of the EDQNM2 closure model with DNS, which validates the results
of the model for the given range of parameters attainable
with DNS.
However, due to the very way it is constructed and implemented,
the added value of the EDQNM model is clearly to allow the investigation
of an extended range of turbulent regimes. In the following two
sections, we investigate high Reynolds number turbulence, currently out of
the grasp of Direct Numerical Simulations (section~\ref{sec:edqnmres}), 
and a derived model for
the limit case of two-dimensional three-components turbulence (section~\ref{sec:2D3C}). 
\subsection{High Reynolds number turbulence}
%---------------------------------------------
\label{sec:edqnmres}
In this section, we address an important question of this article: 
are DNS predictions reliable to understand high Reynolds number quasi-static 
MHD turbulence given the moderate hydrodynamic Reynolds number?
%The previous sections were devoted to the comparison between DNS and EDQNM 
%spectral closures.
%We propose in this section to increase the Reynolds number to values far 
%beyond DNS possibilities.
We use the EDQNM model at higher Reynolds number to answer this question.
The number of wave numbers considered in EDQNM models has to be increased, 
along with the angular discretization and triadic interactions count.
The following simulations are based on $100$ wave numbers, $48$ polar angles 
and $48$ angles for the direction of the plane
of the triad around $\vk$ (denoted $\lambda$ in appendix \ref{appendixA}).
The initial Reynolds number is increased from the previous value of 
$Re\approx 333$ up to $Re\approx 2\times 10^5$.
The initial condition for these high Reynolds simulations is similar to the one 
used for previous EDQNM simulations, except that the inertial range of the initial 
energy spectra is extended to higher wave numbers.
\begin{figure}
\unitlength 0.5mm
\begin{picture}(250,190)
        \put(0,-2){\includegraphics[height=180\unitlength]{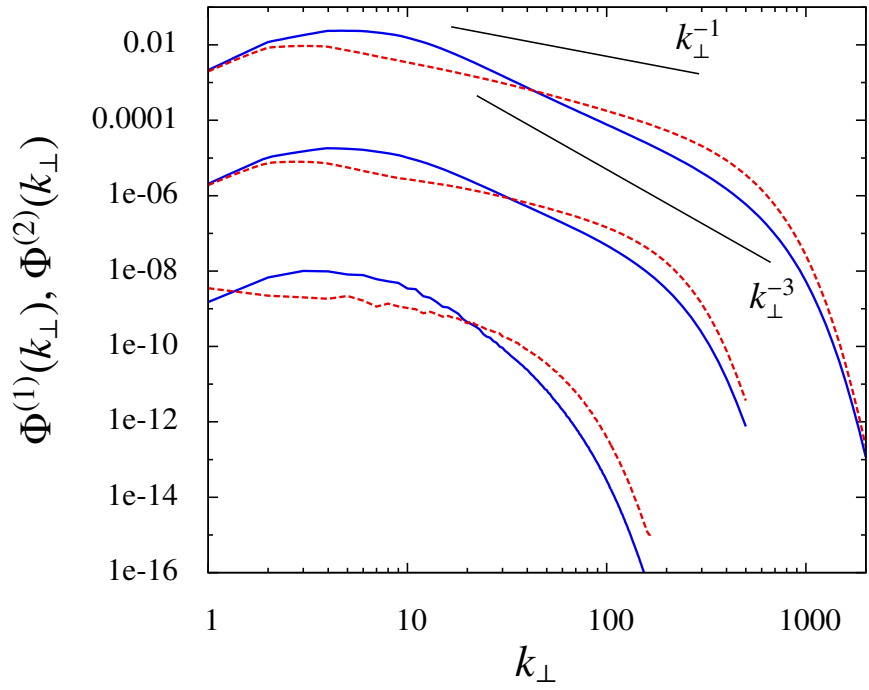}}
\end{picture}
\caption{Poloidal/Toroidal decomposition of the equatorial energy spectra. 
The interaction parameter is $N=5$. All the results are plotted at $t^*\approx5$.
 From bottom to top: DNS using $512^3$ Fourier modes (shifted down by 6 decades),
 EDQNM2 corresponding to a spectral resolution of $1500^3$ Fourier modes 
(shifted down by 3 decades) and EDQNM2 corresponding to a spectral 
resolution of $6000^3$ Fourier modes.\label{fig:highresp4}}
\end{figure}

Equatorial spectra are gathered on figure \ref{fig:highresp4}.
First, the lowermost spectra on the figure recall the previous $512^3$ DNS results 
(these spectra are shifted down by six decades).
The corresponding cross-over wave number  is $k_\perp^c\simeq 171$.
The intermediate results correspond to EDQNM2, obtained with an initial 
 Reynolds  $\textit{Re}\simeq 2200$.
The  DNS resolution required to accurately simulate such a flow is about 
$1500^3$ Fourier modes.
The results are qualitatively unchanged, but the slopes $k_\perp^{-1}$ and 
$k_\perp^{-3}$ appear more clearly, even more so for the top curves on the figure, 
corresponding to an initial 
Reynolds number $\textit{Re}\simeq 2\times 10^5$.
The corresponding DNS resolution using pseudo-spectral methods would be 
about $6000^3$ Fourier modes.
We note also that the cross-over wave number $k^c_\perp$ defined by 
$\Phi^{1}(k^c_\perp)\approx \Phi^{2}(k^c_\perp)$ 
increases with the value of the Reynolds number, to $k^c_\perp\simeq 35$ for 
$\textit{Re}\simeq 2200$, and $k^c_\perp\simeq 50$ for 
$\textit{Re}\simeq 2\times 10^5$.
%
%%%%%%%%%%%%%%%%%%%%%%%%%%%%%%%%%%%%%%%%%%%%%%%%%%%%%%%%%%%%%%%%%%%%%%%%%%%%%%%%%
%
\subsection{A model for 2D-3C turbulence}
%----------------------------------------
\label{sec:2D3C}
In two-dimensional three-components (2D-3C) flows, the velocity
field contains three non zero components, which only vary in two directions 
(the transverse plane, say), and are independent of  the third direction
(axial).
The analogy between 2D-3C turbulence and the final state of quasi-static MHD 
turbulence is supported by theoretical (\cite{mont82}) and numerical 
(\cite{favier10}) evidences.
In previous sections, we found some indications, using EDQNM spectral closures, 
that this statement, supported by DNS at moderate Reynolds number, is valid 
for higher values of the Reynolds number, using EDQNM spectral closures.
This last section is devoted to the comparison between DNS and EDQNM closures 
in a 2D-3C context.

Theoretically, to consider 2D-3C turbulence is equivalent to considering purely 
2D turbulence with a passive scalar (the latter being the vertical component 
of the velocity). As shown by \cite{CAMBON-GODEFERD-1993} (see appendix~\ref{sec:appendix2D3C}),
the EDQNM1 model for anisotropic turbulence reduces exactly to a 2D-3C model for $\Phi^1$
and $\Phi^2$, in which $\Phi^2$ plays the same role as the scalar spectrum
in 2D EDQNM (\cite{HERRING-2D}).

The previous 3D simulations tend to a 2D-3C state but this transition is 
triggered by dissipative effects so that the remaining energy is very small.
To numerically investigate the 2D-3C state at high Reynolds numbers, 
we consider initially 2D-3C turbulence using both a 2D pseudo-spectral code 
and a 2D version of EDQNM closures presented above which include a passive scalar 
(considered here as the axial velocity component).
We use $1024^2$ Fourier modes for the DNS and \textcolor{black}{$51$} wave 
numbers for the spectral discretization of EDQNM.
The initial condition is the same in both cases: 
$\Phi^{1}(k,t^*=0)=\Phi^{2}(k,t^*=0)=10^{-4}k^2\exp(-(k/k_m)^2)$, and $k_m=8$.
The molecular viscosity is fixed to $\nu=5\times 10^{-5}$ which corresponds 
to an initial Reynolds number of about $10^3$.
In the 3D axisymmetric case, the equatorial initial condition was 
also characterised by $\Phi^{1}(k_{\perp})=\Phi^{2}(k_{\perp})$, the main 
difference being that triple correlations were initially non zero.
Here, the initial condition is a random Gaussian velocity field with 
an integral scale $l_0\approx 0.32$ and \textit{rms} velocity $u_0\approx 0.18$,
hence with zero third-order moments.
\begin{figure}
\unitlength 0.5mm
\begin{picture}(300,110)
        \put(-6,-2){\includegraphics[height=100\unitlength]{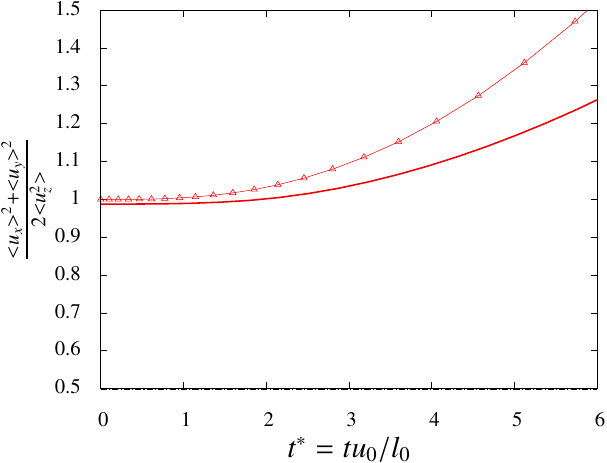}}
        \put(140,-2){\includegraphics[height=100\unitlength]{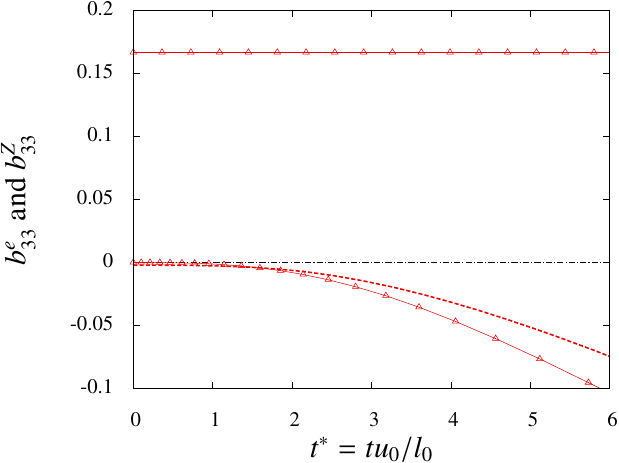}}
        \put(25,16.5){(a)}
        \put(174,18){(b)}
\end{picture}
\caption{\label{fig:2d3c}(a) Ratio between horizontal and vertical kinetic energy for
2D-3C  turbulence. 
(b) Anisotropic tensor $b_{33}$ and its decomposition. --- DNS and $\triangle$ EDQNM.}
\end{figure}
\begin{figure}
\unitlength 0.5mm
\begin{picture}(250,190)
        \put(0,-2){\includegraphics[height=180\unitlength]{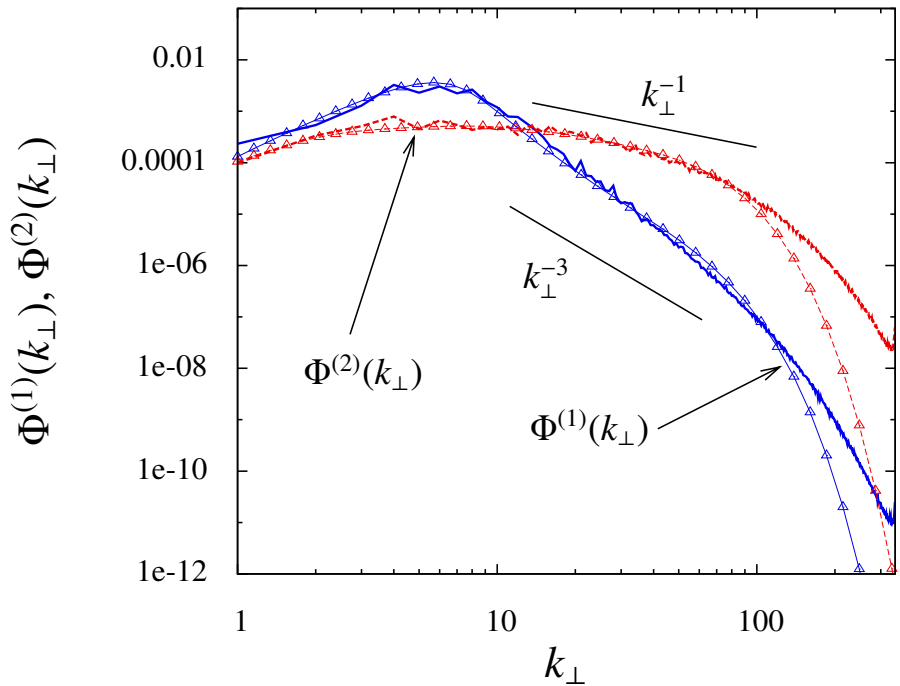}}
\end{picture}
\caption{Poloidal/Toroidal decomposition of the equatorial energy spectra for 
two and a half dimensional turbulence. All the results are plotted at $t^*\approx5$. 
--- DNS, $\triangle$ EDQNM.\label{fig:highrespff}}
\end{figure}

The ratio between transverse and axial kinetic energies is presented in 
figure \ref{fig:2d3c}(a).
As expected, the initial value is about unity.
As time increases, the inverse cascade of the horizontal velocity field develops 
so that the dissipation of horizontal components is reduced.
This phenomenon is responsible for the growth of $\langle u_x^2+u_y^2\rangle/\langle u_z^2\rangle$.
In the 3D axisymmetric case, one first observes a decrease of this quantity 
(see figure \ref{fig:uvw}(a)).
This is due to the transition from a 3D initial state to a quasi-two-dimensional 
state in which the inverse cascade occurs.
Figure~\ref{fig:2d3c}(a) also shows that the EDQNM evolution is faster (maybe
from the fact that, in the EDQNM model, triple correlations---energy transfers---
build up instantly).

On figure \ref{fig:2d3c}(b), the anisotropic tensor $b_{33}$ and its 
$b_{33}^{(e)}$, $b_{33}^{(\mathcal{Z})}$
decomposition 
are presented.
Since the flow is 2D-3C, all axial derivatives are zero, so that all the energy 
is concentrated in the transverse plane.
In that case, $b_{33}^e$ reaches its maximum value $1/6$ (see \cite{CAMBON-MANSOUR-GODEFERD}).
The polarization is initially very small, and becomes negative at larger times 
due to the dominance of toroidal (also transverse) energy with respect to the 
poloidal (also axial) energy.

Finally, the poloidal/toroidal decomposition of the equatorial energy spectra 
is plotted in figure \ref{fig:highrespff}. EDQNM and DNS are in very good agreement
(again with the minor exception of the dissipative range). 
This result confirms the previous $k_\perp^{-3}$ and $k_\perp^{-1}$ 
scalings for the axial and transverse velocity 
components, similar to those observed on figure~\ref{fig:spange} for the
three-dimensional simulations. 
The asymptotic state of quasi-static MHD turbulence is therefore very 
similar to two-dimensional turbulence advecting a passive scalar.
%
%%%%%%%%%%%%%%%%%%%%%%%%%%%%%%%%%%%%%%%%%%%%%%%%%%%%%%%%%%%%%%%%%%%%%%%%%%%%%%%%%%%%%%%
%
\section{Conclusion and final remarks}
%----------------------------------
\label{sec:conclu}
In this paper, we have investigated  the dynamics and the
detailed anisotropy of magneto-hydrodynamic turbulence in the quasi-static
approximation at small magnetic Reynolds number,
 using Direct Numerical Simulations and a two-point statistical
closure of EDQNM type. By essence, such closures consider statistical
averages, which is a key advantage when considering turbulent flows,
for two reasons: first, only one simulation is required to obtain
averaged results, in constrast with the large number of realizations
needed in DNS (typically more than
a dozen); secondly, the obtained averages are smoother
functions than in DNS, all the more if one considers high
order moments (\textit{e.g} third-order correlations).
In terms of computational cost, isotropic EDQNM or the 2D-3C model presented
in section~\ref{sec:2D3C} are thousands of times less costly
than equivalent DNS. The axisymmetric anisotropic EDQNM2 model
abandons one symmetry with respect to the isotropic context, thus
the convolution integral is an order of magnitude more expensive.
Therefore EDQNM2 computations, although not as cpu and memory demanding as
DNS by a factor of about 10 in the present parameter range, are also
run on a parallel computer.  The extension of two-point statistical
closures to bounded turbulent flows (\cite{KRAICHNAN72,TURNER, LAPORTA}), however, 
is analytically and computationally challenging.

In terms of statistical analysis, the closure allows for easy access to the general
decomposition of tensors in the axisymmetric flow, such that refined
statistics of turbulence can be used for characterizing anisotropy. 
The poloidal/toroidal decomposition of the velocity field and related second-order statistics
permits the computation of a polarization tensor, which is a key indicator
of whether the anisotropic mechanism is of linear nature---the Joule dissipation---or
due to more complex nonlinear interactions. (The extraction of equivalent
second-order statistics in physical space, although formally possible, would
be hardly tractable, because of the differential
operators involved). There remains the possibility to obtain such statistics
by post-processing DNS data fields, although with all the inaccuracies and
sub-sampling issues due to limited resolutions. Clearly, DNS discretization
is insufficient in the very large scale range of the spectrum, 
and EDQNM is better off in this
range and very adequate in the inertial range; less so in the 
smallest scales.

We have nonetheless compared results of the EDQNM closure model with
those of $512^3$ DNS. The EDQNM1 version of the model and the EDQNM2 one
provide slightly different results, but the overall agreement with DNS
is quite good. Comparisons involve kinetic energy and enstrophy, kinetic 
energy spectra and directional velocity correlation lengths. The latter
allow us to address the question of numerical confinement due to the finiteness
of the computational box in DNS, of importance in QS MHD turbulence in which
the axial velocity correlation length increases tremendously. 

Several quantities were used to assess the level of anisotropy in
the flow. Starting with initial conditions of isotropic turbulence,
the ratios of transverse energy (resp. enstrophy) to axial
energy (resp. enstrophy), the Shebalin angles and the off-diagonal components
of the Reynolds stress tensor indicate all that the flow dynamics becomes
closer to a two-dimensional three-components state. However, upon 
investigation of transverse and axial energy spectra, we are able to
define a cross-over wavenumber below which the toroidal contribution
dominates over the poloidal one, with a reversal of this order in the larger
wave numbers or small scales. Not only are these predictions of DNS confirmed
in a satisfactory quantitative manner by EDQNM, but the model allows
to reach higher Reynolds numbers than permitted by DNS.
% From WB
%The qualitative trend observed in closure and DNS agree. Furthermore,
The dynamics is not significantly altered at higher Reynolds numbers
reached with the closure model. However, asymptotic scaling behaviour 
appears only very slowly. If a qualitative understanding of QS MHD is called
for, both DNS and closure models are applicable. However, if scaling ranges 
and inertial range behaviour are of interest, two-point closures remain 
an indispensable tool. 
%

% The version of the model has however to be adapted to the case
% in consideration. For instance, rotating turbulence bears strong similarities
% with QS MHD turbulence, although the linear operator
% $M\cos2 \theta$ is of diffusive nature in QS MHD, but its counterpart in
% rotating turbulence is dispersive, generated by the dispersion law of
% inertial waves ($\pm 2 \mathrm{i} \Omega \cos\theta$ with $\Omega$
% the rotation rate). The EDQNM1 model will thus
% be suitable for QS MHD flows, whereas EDQNM2 is required for rotating
% turbulence.
We conclude by noting that rotating turbulence bears strong similarities
with QS MHD turbulence.  
In both cases,   a transition from 3D to 2D structure is observed, and the
2D-2C trend is evidenced
by the separation of $L^{(3)}_{11}$ and $L^{(3)}_{33}$ integral
scales, due to the growth of polarization in the horizontal transverse wave plane.
This transition originates from the \textit{linear} Joule dissipation
term in QS MHD, but from \textit{nonlinear} interactions dominated 
by cubic transfer terms such  as $T^{(e)}$, when solid body rotation acts.
Therefore, QS MHD turbulence may eventually become fully two-dimensional,
whereas complete two-dimensionalization cannot be achieved in rotating turbulence 
in absence of additional phenomena.

\vskip 0.5em
The authors thank the computing centre IDRIS of CNRS for the allocation of 
CPU time under project numbers 071433 and 022206.
We also would like to thank the referees for they suggestions leading
to improvements of the paper. 

\appendix

\section{Detail on anisotropic EDQNM equations and their numerical calculation}
%-----------------------------------------------------------
\label{appendixA}
%

% \textcolor{red}{Quelques ajustements à finir, 
% notamment dans l'introduction des $\mu$, $\eta$, $\theta_{kpq}$...}

\subsection{EDQNM closure for  the spectral energy transfers}
%-------------------------------------
In section~\ref{sec:exact}, the important term to specify is the quasi-normal one
denoted $\Omega^{(QN)}_{s s' s''}(\vk, \vp, t')$, for modelling
the fourth-order terms in~(\ref{cubict}) which is exactly given as a sum of
quadratic terms from the set
\begin{equation}
e = e(\vk, t'), e' =e(\vp, t'), e'' =e(\vq, t'), 
Z = Z(\vk, t')
%, Z'= Z(\vp, t'), Z'' =Z(\vq, t'),
\end{equation}
in the case of a zero helicity flow. (The helicity, in contrast 
with the  polarization
anisotropy, remains zero if initially zero.)
Instead of expressing $\Omega^{(QN)}_{s s' s''}$, it is simpler to derive
its contribution to  $T^{(e)}$ and $T^{(\mathcal{Z})}$, as was
done for the EDQNM model
for rotating turbulence, so that the numerical code for the EDQNM models
used here is easily derived from the one for rotating turbulence
(see \textit{e.g.} \cite{bell06}).
% Recall that the EDQNM3 version was the best one for rotating turbulence,
% with matching to inertial wave turbulence in the asymptotic limit
% of vanishing Rossby number (Bellet {\it et al.} 2006).

%\textcolor{red}{Les \'equations suivantes doivent \^etre donn\'ees directement
%pour le cas QS MHD !}

Detailed equations for $T^{(e)}$ and $T^{(Z)}$ in the EDQNM2 model 
%(resulting from
%the markovianisation described in equation~\ref{eq:mickey}) 
are  
\begin{DIFnomarkup}
$$
T^{(e)}=\frac{1}{2^3}\sum_{s s' s''}\int \frac{2p}{k}
\frac{C^2_{kpq}}{\theta_{kpq}^{-1}
+M_0^2(\cos^2\theta_k +\cos^2\theta_p +\cos^2\theta_q)}\qquad\qquad\qquad\qquad\mbox{}
$$
$$
\left[{A_1(sk, s'p, s'' q)}e''(e -e')
+{A_2(sk, s'p, s'' q)}e^{2\mathrm{i} s''\lambda'' }eZ(s''{\bf q})
+{A_3(sk, s'p, s'' q)}e^{2\mathrm{i} s\lambda }e''Z(s{\bf k})\right.
$$
$$
-{A_5(sk, s'p, s'' q)}e^{2\mathrm{i} s''\lambda'' }e'Z(s''{\bf q})
$$
\begin{equation}
\left.
+{A_4(sk, s'p, s'' q)}\left(e^{2\mathrm{i} s''\lambda'' 
+2\mathrm{i} s\lambda }Z(s''{\bf q})Z(s{\bf k})-e^{2\mathrm{i} s''\lambda'' 
+2\mathrm{i} s'\lambda'}Z(s''{\bf q})Z(s'{\bf p})\right)
\right]
\mathrm{d}^3{\bf p} \label{teomega3}
\end{equation}
and
\[
T^{(z)}=\frac{1}{2^3}\sum_{s' s''}\int \frac{2p}{k}
\frac{C^2_{kpq}e^{-2\mathrm{i}\lambda}}{\theta_{kpq}^{-1}
+M_0^2(\cos^2\theta_k +\cos^2\theta_p +\cos^2\theta_q)}
\qquad\qquad\qquad\qquad\mbox{}
\]
\[
\left[
{A_3(k, -s'p, -s'' q)}e''(e'-e)
+{A_4(k, -s'p, -s'' q)}e^{2\mathrm{i} s''\lambda'' }eZ(s''{\bf q})
+{A_1(k, -s'p, -s'' q)}e^{2\mathrm{i} \lambda }e''Z({\bf k})\right.
\]
\[
-{A_5(k, -s'p, -s'' q)}e^{2\mathrm{i} s'\lambda' }e''Z(s'{\bf p})
\]
\begin{equation}
\left. +{A_2(k, -s'p, -s'' q)}\left(e^{2\mathrm{i} s''\lambda'' +2\mathrm{i} \lambda }Z(s''{\bf q})Z({\bf k})
   - e^{2\mathrm{i} s''\lambda'' +2\mathrm{i} s'\lambda'}Z(s''{\bf q})Z(s'{\bf p})\right)\right] 
\mathrm{d}^3{\bf p}\label{tzomega3} 
\end{equation}
\end{DIFnomarkup}

The geometric factors $A_1$ to $A_5$ are given in the appendix 
of \cite{CAMBON-MANSOUR-GODEFERD}, and in
Sagaut \& Cambon (2008); they depend only on the moduli $k$, $p$, $q$,
`signed' by the polarization signs of helical modes, $s= \pm 1$,
$s' =\pm 1$, $s'' =\pm 1$.  $C_{kpq}$ depends only on the geometry of the 
triad as well, such that
\begin{equation}
\frac{\sin(\widehat {\vp, \vq})}{k} = \frac{\sin(\widehat {\vq, \vk})}{p}
= \frac{\sin(\widehat {\vk, \vp})}{q} = C_{kpq}\ .
\label{ckpq} 
\end{equation}
 The internal triadic
angles $\lambda$, $\lambda'$ and $\lambda''$ denote
the angle of rotation of the plane of the triad around $\vk$, $\vp$, $\vq$,
respectively. Integration variables, which generate all
the other terms at fixed $\vk$, are $p$, $q$, as in
isotropic EDQNM, and $\lambda$, relevant in the axisymmetric case,
and discretized as well.
% Finally, $\sigma_k$ is the unsigned dispersion law of
% inertial wave, or
% \begin{equation} \sigma_k = 2 \Omega \frac{k_{\parallel}}{k}.
% \label{dispomega} \end{equation}  

The only semi-empirical term in the formulae above is the viscous plus
eddy damping term denoted $\theta_{kpq}^{-1}$ since it is homogeneous
to an inverse time scale, with 
\begin{equation} 
\theta_{kpq}^{-1} = \nu(k^2+p^2+q^2)+\vartheta(k,t) + \vartheta(p,t) +\vartheta(q,t), 
\label{mu} 
\end{equation}
in which  $\vartheta(k,t)=A \left(\int^{k}_0 p^2 E(p,t) \mathrm{d}p\right)^{1/2}$ 
may be viewed as a
straining decorrelation time scale of small turbulent structures
by larger ones. Here, $E(k)$ is the classical energy spectrum and 
$A=0.355$
is the only adjusted constant of the model, computed
from the Kolmogorov constant $C_K$ with the 
relation $C_K\simeq 2.76 A^{2/3}$ (\cite{LESIEUR-OSSIA}).

The EDQNM1 version of the closure model does not incorporate
linear Joule dissipation terms proportional to $M^2_0$  in equations~(\ref{teomega3})
and (\ref{tzomega3}). It is therefore generic to any turbulent case,
in which the distorsion only appears explicitly in a linear term added to the dissipation
one.

\subsection{Recovering the 2D-3C case}
%----------------------------------
\label{sec:appendix2D3C}
This was done by \cite{CAMBON-GODEFERD-1993} as follows. In the 2D-3C limit,
$e$ and $Z$  are concentrated in the plane $k_{\parallel}$ (or $k_3$ here)
$=0$, so that 
\begin{equation} e (\vk,t) = e^{(2D)}(k, t) \delta(k_{\parallel}),
Z(\vk, t) = Z^{(2D)} (k,t) \delta(k_{\parallel}),
\end{equation}
and similarly for $T^{(e,Z)}$. 
The Jacobian from $(p_1, p_2)$ to $(p, q)$ variables is now
$1/\sqrt{1-x^2}$, only planar triads ($k_{\parallel}= p_{\parallel} =
q_{\parallel} =0$) are called into play, and 
$e^{2 \mathrm{i} \lambda}= e^{2 \mathrm{i} \lambda'} = e^{2 \mathrm{i} \lambda''} = -1$.
Accordingly, the 2D counterparts of Lin equation for 
$\Phi^1$ and $\Phi^2$ are derived 
%from (\ref{te1}, \ref{tz1}) 
as
\begin{equation}
\left( \frac{\partial }{\partial t} + 2 \nu k^2 \right)\Phi^1(k, t)
= T^1 (k, t) = T^{(e), 2D}(k, t) - T^{(Z), 2D}(k,t)
\end{equation}
and
\begin{equation}
\left( \frac{\partial }{\partial t} + 2 \nu k^2 \right)\Phi^2(k, t)
= T^2 (k, t) = T^{(e), 2D}(k, t) + T^{(Z), 2D}(k, t),
\end{equation}
with
\begin{equation} T^1(k,t) = \int \int_{\Delta_k}
\frac{2kp\theta_{kpq}}{ \sqrt{1 - x^2}} (xy + 2 z^3 -z) \Phi^1(q, t)
\left( \Phi^{1} (p ,t) - \Phi^{1}(k, t) \right) \mathrm{d}p \mathrm{d}q
\end{equation}
and 
\begin{equation} T^2(k,t) = \int \int_{\Delta_k}
\frac{2kp\theta_{kpq}}{ \sqrt{1 - x^2}} (xy + z) \Phi^1(q, t)
\left( \Phi^{2} (p ,t) - \Phi^{2}(k, t) \right) \mathrm{d}p \mathrm{d}q.
\end{equation}

It is shown that the 2D contribution from toroidal (horizontal in 2D)
velocity is governed by the classical isotropic EDQNM equation
restricted to 2D (Leith 1971, Pouquet {\it et al.} 1975), whereas the 2D
contribution from poloidal (vertical in this limit) velocity is governed 
by the isotropic EDQNM equation in 2D for a passive scalar.

More conventional relationship is found in term of the averaged spectrum 
using $e^{(2D)}(k, t) = E(k,t)/
(2 \pi k)$, as for the 3D isotropic case, in which $e(k, t) = E(k,t)/
(4 \pi k^2)$.

\section{RDT solutions for the correlation lengths}
%-------------------------------------------------
\label{appendixB}
The linear inviscid evolution of the spectral tensor is immediately found as
\begin{equation}
e(k, \mu, t) = \frac{E(k, 0)}{4 \pi k^2} \exp\left(-2 M^2_0 \mu^2 t\right), \hspace*{1em}
Z(k, \mu,t) = 0, 
\label{ezrdt}
\end{equation} 
with $\mu= \cos\theta$ and $\theta$ the angle between $\bm{k}$ and the vertical. 

Two-dimensional energy components are invariant when defined as
\begin{equation}
{\langle u^2_3 \rangle }(t) L^{(3) }_{33}(t) = \frac{1}{3} \mathcal{K}_0 l_0, \hspace*{1em}
{\langle u^2_1\rangle }(t) L^{(3)}_{11}(t) = {\langle u^2_2 \rangle }(t) L^{(3)}_{22}(t) = \frac{1}{6} \mathcal{K}_0 l_0,
\label{2decrdt} \end{equation}
because they involve only contributions of $e$ and $Z$ at $\mu=0$.
$\mathcal{K}_0$ and $l_0$ are the initial kinetic energy and initial integral 
scale respectively.
Kinetic energy and individual Reynolds stress components are given by
\begin{equation}
\mathcal{K}(t) = \mathcal{K}_0 \int^1_0 \exp\left(-2 M^2_0 \mu^2 t\right)\textrm{d}\mu \ ,
\label{q2rdt} \end{equation}
and
\begin{equation}
{\langle u^2_3\rangle }(t) =  \frac{\mathcal{K}_0}{2} \int^1_0 (1 -\mu^2) 
\exp\left(-2 M^2_0 t \mu^2 \right)\mathrm{d}\mu, \ {\langle u^2_1\rangle }(t) 
  =  \frac{\mathcal{K}_0}{4} \int^1_0 (1 + \mu^2) \exp\left(-2 M^2_0 t \mu^2 \right)\mathrm{d}\mu,
\label{rstrdt} 
\end{equation}
in agreement with $\mathrm{d}^3 \vk = 2 \pi k^2\mathrm{d}k \mathrm{d}\mu$ using polar-spherical coordinates
for $\vk$ and axisymmetry.

The inviscid RDT time development of all relevant statistical quantities is derived  analytically, in terms of the error function erf (exact relationship available from the authors
upon request). The dominant terms in the evolution yield the following simple scalings:
the kinetic energy decays as $M^{-1}_0 \Gamma (\infty) (2t)^{-1/2}$, as well
as the Reynolds stress components; integral length scales with axial  
separation behave as $M_0 l_0 \sqrt{t}$. 
% Relevant angles, as the Moreau and the Shebalin angles, develop 
% identically, with a law derived from inviscid equation (\ref{eq:tke}), or
% \begin{equation}
% \cos^2 \beta = -\frac{1}{2} \frac{1}{\mathcal{K}} \frac{ d \mathcal {K}}{ d(M^2_0 t)},
% \ \mbox{with} \ \mathcal {K} = \mathcal{K}_0 \frac{1}{M_0 \sqrt{2t}}
% \Gamma (M_0 \sqrt{2t}),
% \label{moreaurdt} \end{equation} 
% with the law for $b^e_{33} = b_{33}$ derived from $b^e_{33} = (1/6) -
% (1/2) \cos^2 \beta$. We note that the enstrophy components behave as
% the Reynolds stress components, because changing $e$ into $k^2 e$ does not
% affect the angular integrals. 
Upon introduction of viscosity through the integrating factor $e^{-2 \nu k^2 t}$
in the integrands of equations~(\ref{q2rdt}) and~(\ref{rstrdt}), the viscous
RDT solution is recovered, this time depending on the explicit shape of the
spectrum $E(k)$. For example, the RDT evolution of integral lengthscales
may be compared to the evolution plotted on figure~\ref{fig:lijlin}, and exhibit
a linear evolution instead as the above inviscid $\sqrt{t}$ behaviour.

% `Viscous RDT' is relevant to discuss, for instance, the results in
% figure \ref{fig:lijlin}. Related solutions are obtained by replacing $\mathcal{K}_0 l_0$ and 
% $\mathcal{K}_0$ in the equations above by their viscous counterpart,
% with a new $e^{-2 \nu k^2 t}$ factor in the integral, 
% so that the result depends on the shape of $E(k,t=0)$, but integrals over
% $\mu$ are unchanged.
% As a consequence, the integral lengthscales in figure  \ref{fig:lijlin} can exhibit a 
% law linear in $t$ instead of a $\sqrt{t}$ one, and enstrophy components
% decay more rapidly than their Reynolds stress counterparts. 
% As an important exception,  the RDT evolution of a non dimensional
% term is not modified (the same with and without 
% viscosity): One recovers the same scaling, in terms of $M^2_0 t$,  for the Moreau angle,  $b^e_{33}$,
% and the Shebalin angle, even if the latter involves the enstrophy spectrum.

\bibliographystyle{jfm}
\bibliography{biblio}

\label{lastpage}

\end{document}